\documentclass[twocolumn]{aastex631}
\usepackage{multirow}
\usepackage{scalerel}
\usepackage{float}
\usepackage{amsmath}	

\begin{document}

\title{Unraveling the kinematic and morphological evolution of the Small Magellanic Cloud}
\shorttitle{Kinematics of the SMC using Gaia DR3}
\shortauthors{Dhanush et al.}

\correspondingauthor{S. R. Dhanush}
\email{srdhanushsr@gmail.com}
\author[0009-0007-0388-3143]{S. R. Dhanush}
\affiliation{Indian Institute of Astrophysics, Bangalore, 560034,  India}
\affiliation{Pondicherry University, R.V. Nagar, Kalapet, 605014, Puducherry, India}
\author[0000-0003-4612-620X]{A. Subramaniam}
\affiliation{Indian Institute of Astrophysics, Bangalore, 560034,  India}
\author[0000-0002-5331-6098]{S. Subramanian}
\affiliation{Indian Institute of Astrophysics, Bangalore, 560034,  India}

\begin{abstract}
 We modeled the kinematics of the Small Magellanic Cloud (SMC) by analyzing the proper motion (PM) from Gaia DR3 of nine different stellar populations, which include young main sequence (MS) stars ($<$ 2 Gyr), red giant branch stars, red clump stars, red giants with line-of-sight velocities, and three groups of star clusters. This analysis was carried out using a robust Markov Chain Monte Carlo method to derive up to 7 kinematic parameters. We trace the evolution from a non-rotating flattened elliptical system as mapped by the old population to a rotating highly stretched disk structure as denoted by the young MS stars and clusters ($<$ 400 Myr). We estimated that the inclination, $i$ ($\sim$ 58$^\circ$ to 82$^\circ$) decreases and the position angle, $\Theta$ ($\sim$ 180$^\circ$ to 240$^\circ$) increases with age. We estimated an asymptotic velocity of $\sim$ 49 - 89 km s$^{-1}$ with scale-radius of $\sim$ 6 - 9 kpc for the young MS populations with velocity dispersion of $\sim$ 11 km s$^{-1}$, suggesting a rotation-supported disk structure. Our models estimate a line-of-sight extension of $\sim$ 30 kpc, in agreement with observations. We identified four regions of the SMC showing anomalies in the residual PM, the East Anomaly (EA), South East Anomaly (SEA), South Anomaly (SA), and West Anomaly (WA). The SEA appears like an infalling feature and is identified for the first time. The tidal imprints observed in the residual PM of the SMC suggest that its evolution is considerably shaped by the recent interaction with the Large Magellanic Cloud.

\end{abstract}

\keywords{(galaxies:) Magellanic Clouds --- galaxies: star clusters: general --- galaxies: kinematics and dynamics --- galaxies: interactions --- galaxies: evolution}

\section{Introduction} \label{sec:intro}

The Magellanic Clouds (MCs) consist of two irregular dwarf galaxies in the Local Group, the Large Magellanic Cloud (LMC) and the Small Magellanic Cloud (SMC). The LMC and the SMC are located at a distance of 49.59$\pm$0.09 kpc \citep{lmc_dit2019Natur.567..200P} and 62.44$\pm$0.47 kpc \citep{smc_dist2020ApJ...904...13G}, respectively. The interactions between the MCs have resulted in the formation of a few unique features in this system. The Magellanic system consists of the MCs along with the Magellanic Stream, a trail of neutral hydrogen that spans more than 100$^\circ$ in the sky \citep{put2003ApJ...586..170P,nid2008ApJ...679..432N,Dngh2016ARA&A..54..363D}, the Magellanic Bridge, which features both stellar and gaseous components \citep{gard10.1093/mnras/266.3.567,mull_bekki_10.1111/j.1745-3933.2007.00356.x}, and the Leading Arm, another gaseous stream with multiple filaments \citep{putman_1998Natur.394..752P,venz2012A&A...547A..12V}. However, the evolution of MCs is influenced not only by their mutual interactions but also by their interactions with the Milky Way \citep{Weinberg_2000,Diaz_2012,2015hammerApJ...813..110H}.

Significant efforts have been made to understand the structure of the SMC. The galaxy is traditionally thought to be a spheroidal or ellipsoidal 3D system, and the overall feature of the SMC is governed by its intrinsic hydrodynamics rather than the tidal interactions \citep{zaristsky2000ApJ...534L..53Z}. According to \cite{cio2000A&A...358L...9C}, younger stars in the SMC display an irregular structure characterized by spiral arms and tidal features. In contrast, the studies on the older population indicate that they are being distributed in a spherical or ellipsoidal manner (\citealt{3dsm_as2012ApJ...744..128S}, hereafter, S12; \citealt{deb8209373}; \citealt{spheroidal_rubelo_}, hereafter, R18; \citealt{yousafi2019MNRAS.490.1076E}, hereafter, Y19). Studies by \cite{her2012AJ....144..107H}, \citealt{smith2015A&A...573A.135S} (hereafter S15), and \citealt{Teodoro_10.1093/mnras/sty3095} (hereafter D18), suggested that the SMC can have a disk morphology as well. \cite{Deb_2019MNRAS.489.3725D} showed that the northeastern part of the SMC bar is closer to us than the southwestern part. The bar is notably more pronounced in the youngest main sequence stars of the SMC and appears fragmented, potentially as a result of tidal interactions (Y19). 

The estimated line-of-sight (LOS) depth of the SMC is found to be greater than that of the LMC and reaches more than $\sim$ 20 kpcs (S12; \citealt{jacy2016AcA....66..149J}, hereafter, J16; \citealt{jacy2017AcA....67....1J}; \citealt{ripepi10.1093/mnras/stx2096}, hereafter, R17; \citealt{mura10.1093/mnras/stx2514}). The studies of the Red Clump (RC) population have shown a distance bimodality in the SMC as well, suggesting that an eastern foreground stellar population is situated more than $\sim$ 10 kpc in front of the main body of the SMC (\citealt{Nidever_2010}; \citealt{smitha201710.1093/mnras/stx205}; \citealt{yous202110.1093/mnras/stab1075}; \citealt{abhinaya2021MNRAS.500.2757O}; \citealt{dizna10.1093/mnras/stab2873}). 
Recently, \cite{sakowska_202410.1093/mnras/stae1766} reported that the northeastern region of the SMC has a LOS depth of $\sim$ 7 kpc.

Over the past decade, several kinematic studies and models of the SMC have been conducted (\citealt{Costa_2011}, hereafter, C11; \citealt{Kallivayalil_2013}, hereafter, K13; \citealt{vanderMarel_2016}, hereafter, V16; 
\citealt{Zivick_2018}, hereafter, Z18; \citealt{gaia2018A&A...616A..12G}, hereafter, G18; \citealt{deleo}, hereafter, D20; \citealt{Florian}, hereafter, N21). Also, it has been observed that the HI gas of the SMC exhibits considerable internal rotation (\citealt{h1_2_sta2004ApJ...604..176S}, hereafter, S04; D18), while the older population has a weak or little rotation (\citealt{Harris_2006}; Z18; \citealt{Zivick_2021}, hereafter, Z21). The young stars show an ordered motion toward the Magellanic Bridge with a larger proper motion (PM) than that of the main body of the SMC \citep{Oey_2018}. 
 
Previous studies indicate that the distribution of younger and older populations within the SMC differs significantly. Additionally, the morphology and kinematics of the SMC vary depending on the specific tracer population used in the study. Moreover, many earlier studies do not adequately address the estimated LOS depth using a model. To comprehend the substantial LOS depth, morphology and structure of the SMC, it is essential to model the kinematics of the galaxy as a function of different populations (both young and old). In our study, we address these aspects by presenting a 2D model of the SMC that incorporates various populations, both young and old, utilizing data from \textit{Gaia} DR3. Additionally, we develop a 3D model of the SMC using Red Giants that have LOS velocity (v$_{los}$) information as well.

In this study, we aim to achieve the following: (1) to develop a base kinematic disk model of the SMC that aligns with the observed kinematics of both young and old stars in the galaxy; (2) to detect the signals of rotation in the SMC, based on the age of the stellar population; (3) to estimate the kinematic disk parameters, which aid in revealing the orientation, galaxy plane morphology, and LOS features of the SMC; (4) to investigate the tidal interaction/influence on the SMC by the LMC, based on the kinematic response of both young and old population. Addressing these points will offer insights into the structural and kinematic evolution of the SMC within the Magellanic system.

The paper is arranged as follows. In Section \ref{sec:data}, we describe the datasets used for our modeling. In Section \ref{sec:model}, we describe the procedures of our kinematic model. In Section \ref{sec:results}, we present the estimated kinematic parameters of our datasets, followed by the estimation of the residual PM. The discussions based on our results are presented in Section \ref{sec:disc}, followed by Section \ref{sec:summary}, which summarizes our work.

\section{Data} \label{sec:data}

\begin{figure}
    \centering
       \includegraphics[width=1\linewidth]{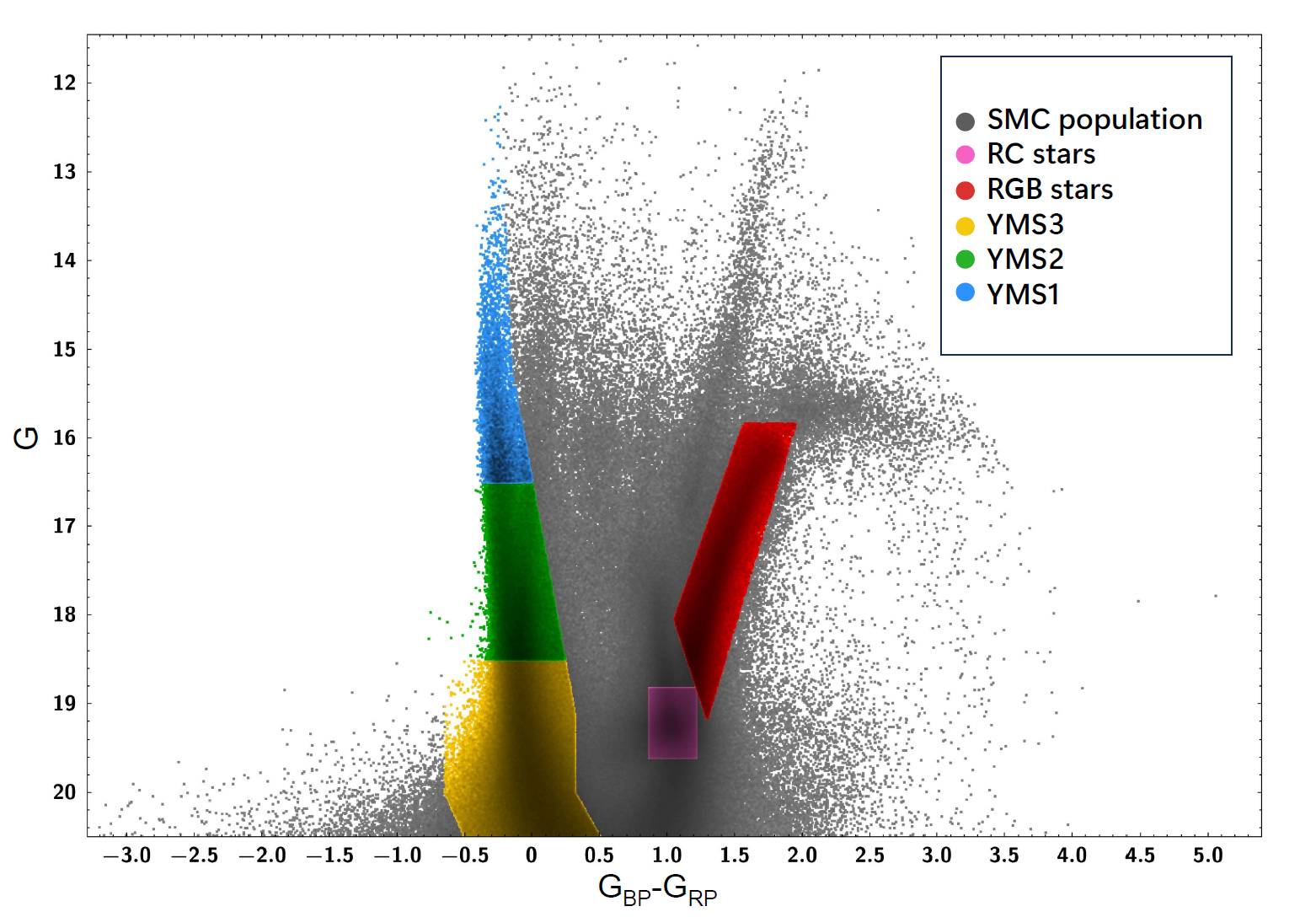}
       \caption{The CMD illustrating the selection of various stellar populations in the SMC as defined in section 2 is presented here. The populations are highlighted in corresponding colors: RC stars (pink), RGB stars (red), YMS3 (yellow), YMS2 (green), and YMS1 (light blue).}
       \label{fig:cmdselection} 
\end{figure}

To model the young and old population of the SMC, we used the cleaned SMC \textit{Gaia} DR3 catalog provided in the study by \citet[hereafter, J23]{Jimsmc2023A&A...672A..65J}, which is produced based on a robust neural network classifier. We retained the sources with Probability, P $\geq$ 0.31 for optimal selection of the SMC sources, as recommended by the authors. We selected the SMC sources within a box of 7$^{\circ}$.5 centered on the SMC. We adopted the optical center of the SMC ($\alpha_0=$ 13$^\circ$.05 and $\delta_0=$ -72$^\circ$.83, \citealt{DEVAUCOULEURS1972163}) in our modeling. We then classified the SMC sources into sub-samples as Young Main Sequence 1 (YMS1, ages $<$ 50 Myr), Young Main Sequence 2 (YMS2, 50 $<$ age $<$  400), Young Main Sequence 3 (YMS3, mixed ages reaching up 1-2 Gyr), and Red Giant Branch (RGB), based on the polygon selections in the Color Magnitude Diagram (CMD) of the SMC as provided in Section 2.3.2 of \citet[hereafter, G21]{gaia2021A&A...649A...7G}. We also selected the Red Clump (RC) population from the CMD based on the selection criterion mentioned in \cite{saroo2022A&A...666A.103S}. The selected sub-samples are highlighted and displayed on the CMD, as shown in Figure \ref{fig:cmdselection}. Since these datasets lack LOS velocity information, we cross-matched the J23 catalog with the Red Giants catalog by \citet[hereafter, D14]{Do2014MNRAS.442.1663D} and found 3545 stars, which have LOS velocity information as well. 

We also used the dataset of 280 star clusters in the SMC from the analysis of \citet[hereafter, D24]{srd_paper1}, spanning ages from $\sim$ 12 Myr to 3.4 Gyr, that used Gaia DR3 to estimate the age, metallicity, extinction, and distance modulus. The clusters were parameterized after field star decontamination, as explained in section 2 of D24. The clusters were divided into three age groups: (1) Young clusters (CLSY - ages $<$ 400 Myr, 143 clusters), (2) Intermediate age clusters  (CLSI - 400 Myr $<$ ages $<$ 1 Gyr, 65 clusters), and (3) Old clusters (CLSO - ages $>$ 1 Gyr, 72 clusters).

The sources shown in Figure \ref{fig:cmdselection} are then orthographically projected with the SMC center, using the equations outlined in G21. We used a bin size of 0$^\circ$.25 to bin the sources and calculated the median PM and their corresponding standard errors for each bin. We retained bins with more than 5 stars, which resulted in 140 YMS1, 372 YMS2, 1093 YMS3, 1288 RGB, and 1731 RC bins as the final datasets for the modeling. The bin size and minimum number of stars per bin are determined through multiple trials with different bin sizes and star counts. The optimal combination is selected based on the visual clarity of features observed in the vector point diagram (VPD) of bulk PM and residual PM (see Section \ref{sec:results}). The radial coverage of Red Giants from D14 is mostly within $\sim$ 3$^\circ$ of the SMC, so we model them as individual sources rather than binning the data. For clusters, we calculated the median PM and the associated standard error of stars within each cluster.

The datasets mentioned above are used for modeling the SMC. The following section provides the analytical background and the modeling procedure.

\section{Kinematic model of the SMC}\label{sec:model}

The following subsections offer an overview of the theoretical framework used in our modeling, as well as the Bayesian methodology employed to estimate the optimal kinematic parameters for the nine datasets.

\subsection{Analytical background of the modeling}

We performed the kinematic model of the SMC based on the equations outlined by \citet[hereafter, V02]{van2002AJ....124.2639V}. The method involves defining the directional vectors of local PM of sources in the West and North (M$_W$, M$_N$) in the sky plane using a series of several independent model parameters. The parameters selected for our model encompass the inclination of the SMC disk ($i$), the position angle of the line of nodes measured from West ($\theta$), the amplitude of the tangential velocity of the SMC's center of mass ($v_t$), the tangential angle made by $v_t$ ($\theta_t$), scale radius ($R_f$), the asymptotic velocity ($v_f$), and the systemic velocity ($v_{sys}$). Our modeling process is aimed at determining the best-fitting values for these kinematic parameters. It should be noted that the disk we expect to trace is the projected distribution of the ellipsoidal SMC along its semi-major and semi-minor axes.

We assumed that the PM of sources in the SMC includes both the center of mass (COM) motion and rotational components within the disk model. We used the parametric equation for rotation as described in D18 and also assumed that the SMC disk experiences no precession or nutation.

The following subsection describes the modeling procedure employed to estimate the best-fitting kinematic parameters for the datasets discussed in Section \ref{sec:data}.

\subsection{Modeling procedure}

The best fitting kinematic parameters of the SMC were estimated using the Markov Chain Monte Carlo (MCMC) serial stretch move sampling algorithm introduced by \cite{gdman2010CAMCS...5...65G}. The implementation of MCMC is similar to the previous study by \cite{Dhanush_2024_kinemtics} on the kinematic model of the LMC. The model PM ($\mu_{W,m}$, $\mu_{N,m}$) and the observed PM ($\mu_{W,o}$, $\mu_{N,o}$) are used to construct the log-likelihood function ($\ln\mathcal{L}$), which can be used with the associated West and North direction standard errors of observed datasets ($\sigma_{W,o}$,$\sigma_{N,o}$) to sample the best fitting parameter with MCMC. For Red Giants with observed LOS velocities (v$_{los,o}$) and associated errors ($\sigma_{los,o}$), we include the model LOS velocity (v$_{los,m}$) in the $\ln\mathcal{L}$ as well.   
The equation for $\ln\mathcal{L}$ is given by,
\begin{align}
    \ln\mathcal{L} &= -0.5 \sum_{i=1}^{n} \left[ \ln \left( 2\pi\sigma_{W,o,i}^2 \right) + \frac{ \left( \mu_{W,o,i} - \mu_{W,m,i} \right)^2 }{\sigma_{W,o,i}^2} \right. \nonumber \\
    &\quad + \ln \left( 2\pi\sigma_{N,o,i}^2 \right) + \left. \frac{ \left( \mu_{N,o,i} - \mu_{N,m,i} \right)^2 }{\sigma_{N,o,i}^2} \right. \nonumber\\\ 
    &\quad + \ln \left( 2\pi\sigma_{los,o,i}^2 \right) + \left. \frac{ \left( v_{los,o,i} - v_{los,m,i} \right)^2 }{\sigma_{los,o,i}^2} \right] 
    \label{eqn:likeli}
\end{align}
It is to be noted that the last two terms in equation \ref{eqn:likeli} are excluded for the eight datasets that lack v$_{los}$ information. Additionally, we consider two primary variants of the model: one that includes the rotation component of the observed PMs and one that does not. Both variants are tested across each dataset to determine which best represents the different stellar populations of the SMC. The final model for each population is chosen based on the variant (with or without rotation) that shows convergence in the posterior distribution of the sampled parameters. For datasets lacking v$_{los,o}$, the $v_{sys}$ is fixed at 145.6 km s$^{-1}$ (V02), but it is treated as a free parameter otherwise. The kinematic center is fixed at $\alpha_0=$ 13$^\circ$.05 and $\delta_0=$ -72$^\circ$.83, which is the optical center adopted to select the coverage of the SMC in this study (Section \ref{sec:data}). The priors for the $i$, $\theta$, $v_t$, and $v_f$ were uniformly chosen, while $R_f$ was assigned a Gaussian prior centered at 1.1 kpc (D18), with an extended 3$\sigma$ range to effectively explore the parameter space. 

We executed 2000 steps of MCMC iteration involving 200 walkers evolving sequentially at each step. From the sampled posterior values for the parameters, we focused on the final 50\%, calculating their median alongside the 16$^{th}$ and 84$^{th}$ percentile errors for parameter estimation. 
In the following section, we present the results obtained from the above modeling procedures.

\section{Results}\label{sec:results}

\begin{figure*}
    \centering
       \includegraphics[width=1\linewidth]{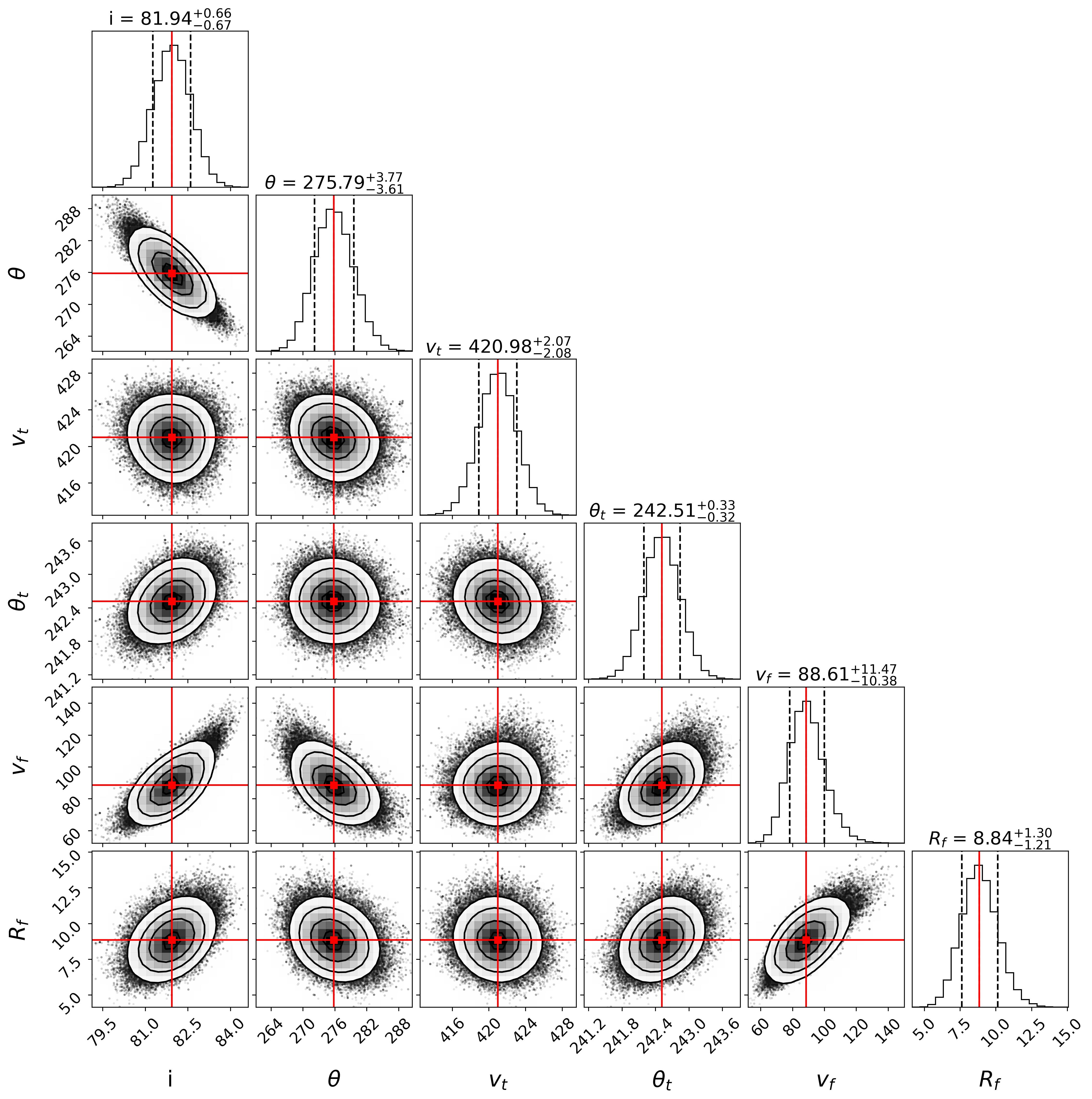}
       \caption{The corner plot representing the sampled posterior distribution of kinematic parameters for the YMS1 is shown here. The vertical red lines represent the median values, and the black dashed lines represent the 16th and 84th percentiles.}
       \label{fig:mcmc_yms1} 
\end{figure*}

\begin{figure*}
    \centering
       \includegraphics[width=1\linewidth]{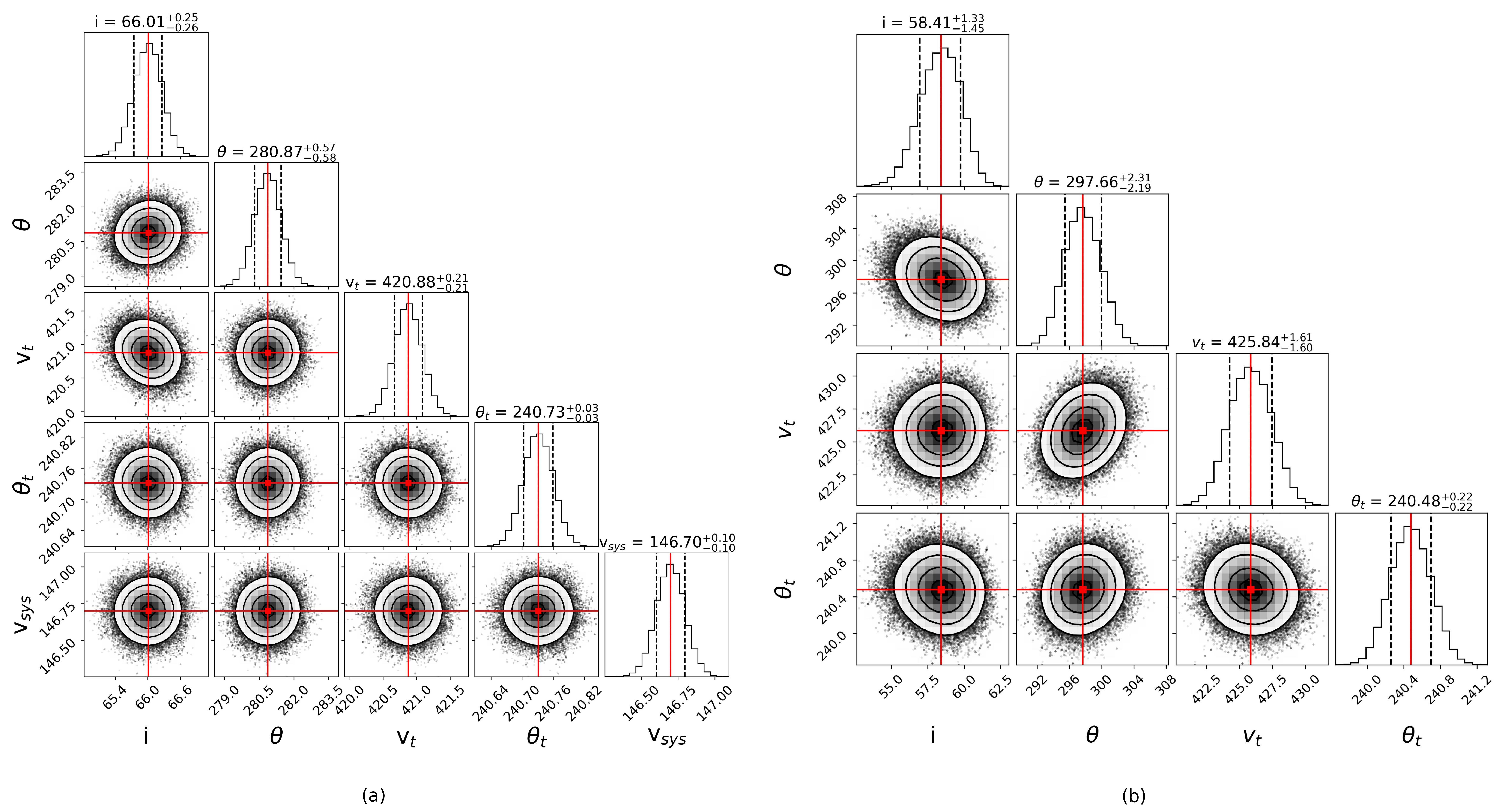}
       \caption{The corner plot representing the sampled posterior distribution of kinematic parameters for the Red Giants with v$_{los}$ (panel a) and RGB without v$_{los}$ (panel b) are shown here. The vertical red lines represent the median values, and the black dashed lines represent the 16th and 84th percentiles.}
       \label{fig:mcmc_rgb_rgb} 
\end{figure*}

In this study, we present a 2D kinematic model of the SMC by analyzing various SMC populations using \textit{Gaia} DR3 data. These populations include YMS1, YMS2, YMS3, RC, RGB, CLSY, CLSI, and CLSO. The study covers a radius within $\sim$ 7$^\circ$.5 from the SMC center. Additionally, we develop a 3D disk model of the SMC using Red Giants and their LOS velocities estimated by D14. The modeling indicates that rotation in the SMC is present in the YMS1, YMS2, CLSY, and CLSI. In contrast, CLSO, YMS3, RGB, RC, and Red Giants with $v_{los}$ in the SMC, show no evidence of significant rotation. Figure \ref{fig:mcmc_yms1} shows the sampled posterior distribution of the kinematic parameters for the YMS1. Figure \ref{fig:mcmc_rgb_rgb} presents a comparison of the sampled posterior distributions of the kinematic parameters for the RGB population without $v_{los}$ and the Red Giants with $v_{los}$. The best-fitting kinematic parameters for all the datasets considered in this study are tabulated in Table \ref{tab1_paraemters}. 

\begin{table*}
    \centering

    \caption{The estimated kinematic parameters for nine populations in our study are provided here. The position angle of the line of nodes is measured from the north ($\Theta$ = $\theta-90^\circ$).}
    \label{tab1_paraemters}
\renewcommand{\arraystretch}{1.5}
    \begin{tabular}{l@{\hspace{1.2em}}c@{\hspace{1.2em}}c@{\hspace{1.2em}}c@{\hspace{1.2em}}c@{\hspace{1.2em}}c@{\hspace{1.2em}}c@{\hspace{1.2em}}c@{\hspace{1.2em}}}
        \hline\hline
        \textbf{Data} & \textbf{$i$} & \textbf{$\Theta$} & \textbf{$v_t$}& \textbf{$\theta_t$}& \textbf{$v_f$}& \textbf{$R_f$}& \textbf{$v_{sys}$} \\
         & (deg) & (deg) & (km s$^{-1}$) & (deg) & (km s$^{-1}$) & (kpc)& (km s$^{-1}$) \\
        \hline
       YMS1& 81.94$^{\hspace{0.01cm}\scaleto{+0.66}{4.7pt}}_{\hspace{0.01cm}\scaleto{-0.67}{4.7pt}}$ &  185.79$^{\hspace{0.01cm}\scaleto{+3.77}{4.7pt}}_{\hspace{0.01cm}\scaleto{-3.61}{4.7pt}}$ & 420.98$^{\hspace{0.01cm}\scaleto{+2.07}{4.7pt}}_{\hspace{0.01cm}\scaleto{-2.08}{4.7pt}}$ & 242.51$^{\hspace{0.01cm}\scaleto{+0.33}{4.7pt}}_{\hspace{0.01cm}\scaleto{-0.32}{4.7pt}}$ & 88.61$^{\hspace{0.01cm}\scaleto{+11.47}{4.7pt}}_{\hspace{0.01cm}\scaleto{-10.38}{4.7pt}}$ &  8.84$^{\hspace{0.01cm}\scaleto{+1.30}{4.7pt}}_{\hspace{0.01cm}\scaleto{-1.21}{4.7pt}}$  & fixed\\
         \hline
        YMS2& 77.68$^{\hspace{0.01cm}\scaleto{+0.90}{4.7pt}}_{\hspace{0.01cm}\scaleto{-0.91}{4.7pt}}$ &  188.92$^{\hspace{0.01cm}\scaleto{+4.19}{4.7pt}}_{\hspace{0.01cm}\scaleto{-4.17}{4.7pt}}$ & 434.21$^{\hspace{0.01cm}\scaleto{+3.04}{4.7pt}}_{\hspace{0.01cm}\scaleto{-2.92}{4.7pt}}$ & 240.97$^{\hspace{0.01cm}\scaleto{+0.40}{4.7pt}}_{\hspace{0.01cm}\scaleto{-0.39}{4.7pt}}$ & 49.32$^{\hspace{0.01cm}\scaleto{+10.91}{4.7pt}}_{\hspace{0.01cm}\scaleto{-9.44}{4.7pt}}$ &  5.85$^{\hspace{0.01cm}\scaleto{+1.65}{4.7pt}}_{\hspace{0.01cm}\scaleto{-1.52}{4.7pt}}$ & fixed \\
         \hline
         YMS3& 71.34$^{\hspace{0.01cm}\scaleto{+1.38}{4.7pt}}_{\hspace{0.01cm}\scaleto{-1.64}{4.7pt}}$ &  200.29$^{\hspace{0.01cm}\scaleto{+4.55}{4.7pt}}_{\hspace{0.01cm}\scaleto{-4.45}{4.7pt}}$ & 447.52$^{\hspace{0.01cm}\scaleto{+5.37}{4.7pt}}_{\hspace{0.01cm}\scaleto{-5.53}{4.7pt}}$ & 241.20$^{\hspace{0.01cm}\scaleto{+0.59}{4.7pt}}_{\hspace{0.01cm}\scaleto{-0.59}{4.7pt}}$ & 
         -& 
         - & fixed \\
         \hline
         CLSY& 79.93$^{\hspace{0.01cm}\scaleto{+0.78}{4.7pt}}_{\hspace{0.01cm}\scaleto{-0.80}{4.7pt}}$ &  180.46$^{\hspace{0.01cm}\scaleto{+3.44}{4.7pt}}_{\hspace{0.01cm}\scaleto{-3.32}{4.7pt}}$ & 422.80$^{\hspace{0.01cm}\scaleto{+1.39}{4.7pt}}_{\hspace{0.01cm}\scaleto{-1.40}{4.7pt}}$ & 241.26$^{\hspace{0.01cm}\scaleto{+0.20}{4.7pt}}_{\hspace{0.01cm}\scaleto{-0.19}{4.7pt}}$ & 
         66.59$^{\hspace{0.01cm}\scaleto{+11.28}{4.7pt}}_{\hspace{0.01cm}\scaleto{-10.23}{4.7pt}}$  & 
         7.12$^{\hspace{0.01cm}\scaleto{+1.40}{4.7pt}}_{\hspace{0.01cm}\scaleto{-1.31}{4.7pt}}$ & fixed\\
         \hline
         CLSI& 74.80$^{\hspace{0.01cm}\scaleto{+1.46}{4.7pt}}_{\hspace{0.01cm}\scaleto{-1.73}{4.7pt}}$ &  182.14$^{\hspace{0.01cm}\scaleto{+5.44}{4.7pt}}_{\hspace{0.01cm}\scaleto{-5.33}{4.7pt}}$ & 420.80$^{\hspace{0.01cm}\scaleto{+2.32}{4.7pt}}_{\hspace{0.01cm}\scaleto{-2.30}{4.7pt}}$ & 241.28$^{\hspace{0.01cm}\scaleto{+0.31}{4.7pt}}_{\hspace{0.01cm}\scaleto{-0.30}{4.7pt}}$ & 
         49.73$^{\hspace{0.01cm}\scaleto{+13.69}{4.7pt}}_{\hspace{0.01cm}\scaleto{-12.18}{4.7pt}}$  & 
         6.33$^{\hspace{0.01cm}\scaleto{+1.86}{4.7pt}}_{\hspace{0.01cm}\scaleto{-1.72}{4.7pt}}$ & fixed\\
         \hline
         CLSO& 63.30$^{\hspace{0.01cm}\scaleto{+2.35}{4.7pt}}_{\hspace{0.01cm}\scaleto{-2.78}{4.7pt}}$ &  240.44$^{\hspace{0.01cm}\scaleto{+4.65}{4.7pt}}_{\hspace{0.01cm}\scaleto{-5.16}{4.7pt}}$ & 419.58$^{\hspace{0.01cm}\scaleto{+1.88}{4.7pt}}_{\hspace{0.01cm}\scaleto{-1.94}{4.7pt}}$ & 240.65$^{\hspace{0.01cm}\scaleto{+0.24}{4.7pt}}_{\hspace{0.01cm}\scaleto{-0.23}{4.7pt}}$ & 
         - & 
         - & fixed\\
         \hline
          RGB& 58.41$^{\hspace{0.01cm}\scaleto{+1.33}{4.7pt}}_{\hspace{0.01cm}\scaleto{-1.45}{4.7pt}}$ &  207.66$^{\hspace{0.01cm}\scaleto{+2.31}{4.7pt}}_{\hspace{0.01cm}\scaleto{-2.19}{4.7pt}}$ & 425.84$^{\hspace{0.01cm}\scaleto{+1.61}{4.7pt}}_{\hspace{0.01cm}\scaleto{-1.60}{4.7pt}}$ & 240.48$^{\hspace{0.01cm}\scaleto{+0.22}{4.7pt}}_{\hspace{0.01cm}\scaleto{-0.22}{4.7pt}}$ & 
         - & 
         - & fixed \\
          \hline
          RC& 58.20$^{\hspace{0.01cm}\scaleto{+1.51}{4.7pt}}_{\hspace{0.01cm}\scaleto{-1.68}{4.7pt}}$ &  202.00$^{\hspace{0.01cm}\scaleto{+3.22}{4.7pt}}_{\hspace{0.01cm}\scaleto{-3.07}{4.7pt}}$ & 421.91$^{\hspace{0.01cm}\scaleto{+2.31}{4.7pt}}_{\hspace{0.01cm}\scaleto{-2.28}{4.7pt}}$ & 241.18$^{\hspace{0.01cm}\scaleto{+0.32}{4.7pt}}_{\hspace{0.01cm}\scaleto{-0.32}{4.7pt}}$ & 
         - & 
         - & fixed \\
         \hline
          Red Giants (D14)& 66.01$^{\hspace{0.01cm}\scaleto{+0.25}{4.7pt}}_{\hspace{0.01cm}\scaleto{-0.26}{4.7pt}}$ &  190.87$^{\hspace{0.01cm}\scaleto{+0.57}{4.7pt}}_{\hspace{0.01cm}\scaleto{-0.58}{4.7pt}}$ & 420.88$^{\hspace{0.01cm}\scaleto{+0.21}{4.7pt}}_{\hspace{0.01cm}\scaleto{-0.21}{4.7pt}}$ & 240.73$^{\hspace{0.01cm}\scaleto{+0.03}{4.7pt}}_{\hspace{0.01cm}\scaleto{-0.03}{4.7pt}}$ & 
         - & 
         - & 146.70$^{\hspace{0.01cm}\scaleto{+0.10}{4.7pt}}_{\hspace{0.01cm}\scaleto{-0.10}{4.7pt}}$ \\
        \hline\hline
    \end{tabular}
\end{table*}

In the following subsections, we detail the results from our models, corresponding to different SMC populations. We highlight some notable features in each population and also present the residual PM map and its distribution.

\subsection{YMS1 and YMS2}\label{yms1_2_results}

\begin{figure*}
    \centering
       \includegraphics[width=1\linewidth]{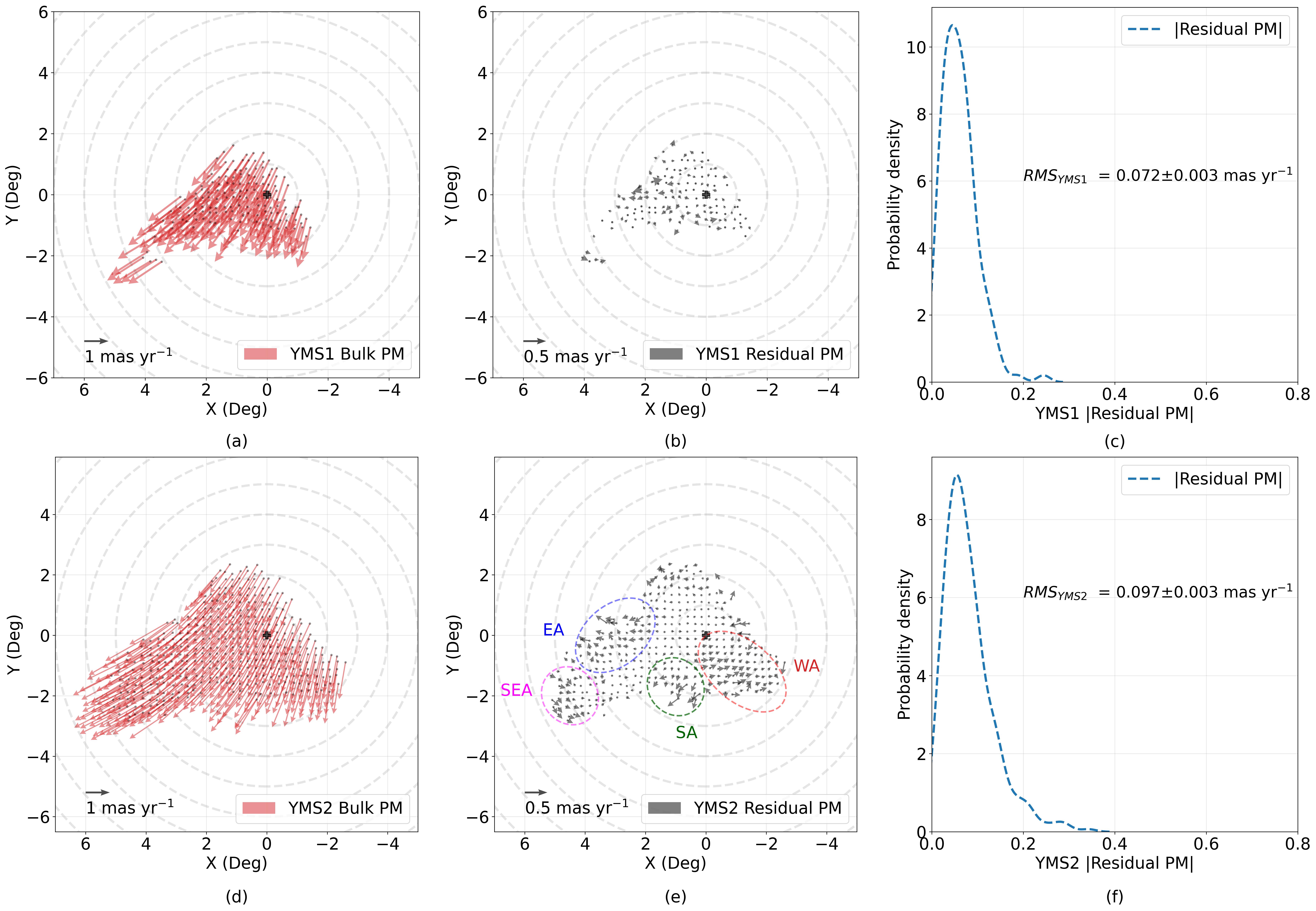}
       \caption{The observed PM (panels a and d), residual PM vectors (panels b and e), and distribution of $|$residual PM$|$ (panels c and f) for the YMS1\&2 are shown here. The EA, SEA, SA, and WA regions identified in the residual PM map of YMS2 are highlighted.}
       \label{fig:yms12_pm} 
\end{figure*}

The observed bulk PM of the YMS1 and YMS2 in the skyplane are shown in Figure \ref{fig:yms12_pm}, panels (a) and (d), respectively. We found these populations support rotation components in their bulk PM. However, we estimated a larger $v_f$ and $R_f$ for YMS1 ($\sim$ 89 km s$^{-1}$ and 9 kpc, respectively) compared to the YMS2 population ($\sim$ 49 km s$^{-1}$ and 6 kpc, respectively). Meanwhile, the estimated COM PM values in the west and north directions ($\mu_{W,com}$, $\mu_{N,com}$) for YMS1 (-0.657$\pm$0.008, -1.262$\pm$0.007) and YMS2 (-0.712$\pm$0.010, -1.283$\pm$0.010) do not show any significant deviation. The estimated $i$ shows an offset of 4$^\circ$.26$\pm$$1^\circ$.13, while the position angle of the line of nodes, measured from north $\Theta$ ($\theta-90^\circ$) does not show any significant variation.

Panels (b) and (e) in the Figure \ref{fig:yms12_pm} depict the residual PM (Observed PM - Model PM) for YMS1 and YMS2, respectively. We visually detected four anomalies in the residual PM map of YMS2 by identifying large deviations in the amplitude and direction of the residual vectors. These include the East Anomaly (EA), which is significant beyond $\sim$ 2$^\circ$ with residual PM directed towards east and northeast; the South East Anomaly (SEA), located in the southeast and extending beyond $\sim$ 4$^\circ$, where residual PM is directed towards south with indication for counter rotation; the South Anomaly (SA), found in the southern region beyond $\sim$ 1$^\circ$, where residual PM is directed towards the southeast; and the more prominent West Anomaly (WA), which extends from the central regions of the SMC to the southwest, with residual PM directed westward. YMS1 exhibits significantly less pronounced SEA, WA, and SA, though the EA is still evident. 

The magnitude of the residual PM distributions ($|$residual PM$|$) for YMS1 and YMS2 are presented in panels (c) and (f), respectively. The Root Mean Square (RMS) values are estimated as 0.072$\pm$0.003 mas yr$^{-1}$ for YMS1 and 0.097$\pm$0.003 mas yr$^{-1}$ for YMS2, respectively.

\subsection{YMS3}
\begin{figure*}
    \centering
       \includegraphics[width=1\linewidth]{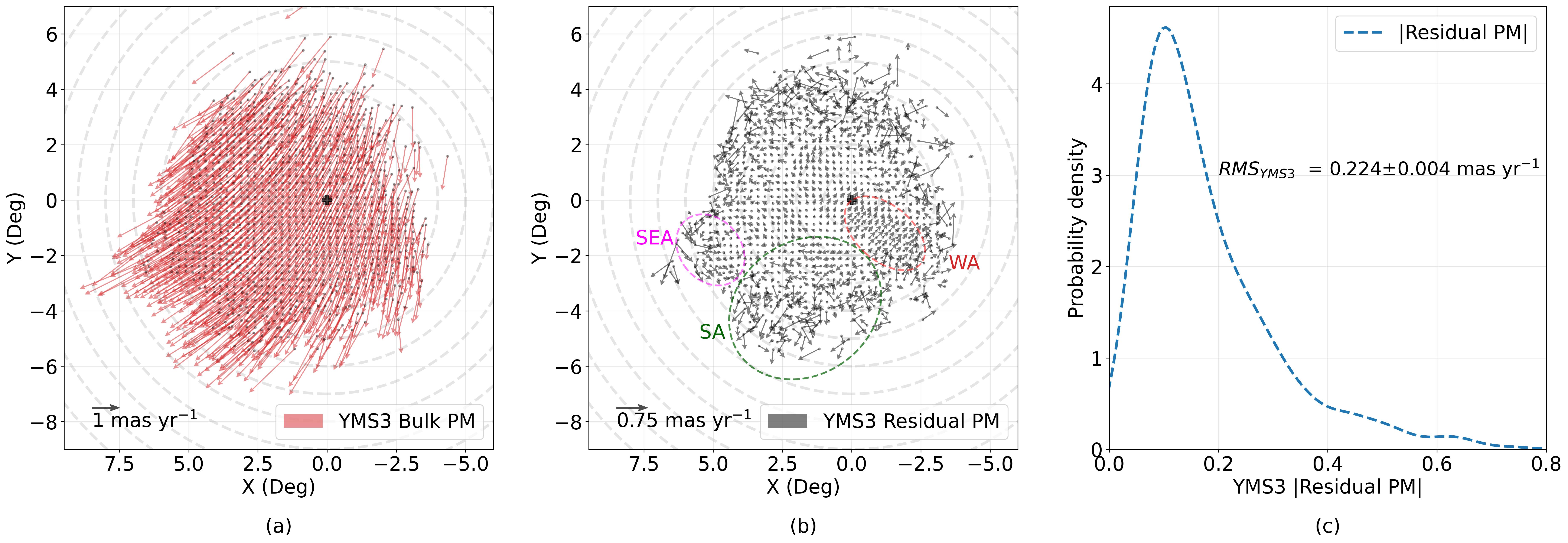}
       \caption{The observed PM (panel a), residual PM vectors (panel b), and distribution of $|$residual PM$|$ (panel c) for the YMS3 are shown here. The SEA, SA, and WA regions identified in the residual PM map of YMS3 are highlighted.}
       \label{fig:yms3_pm} 
\end{figure*}
Figure \ref{fig:yms3_pm}, panel (a) shows the observed bulk PM of YMS3. The spatial distribution of stars in YMS3 is more spatially spread than YMS1 and YMS2. This population does not exhibit significant rotation in their bulk PM, as the model variant that includes rotation fails to achieve convergence in the MCMC sampler. We estimated a slightly larger ($\mu_{W,com}$, $\mu_{N,com}$) for YMS3 (-0.728$\pm$0.016, -1.325$\pm$0.018) when compared to YMS1\&2, notably in the south direction. Also, the estimated $i$ ($\sim$ 71$^\circ$) for YMS3 is lower, and $\Theta$ ($\sim$ 200$^\circ$) is slightly larger when compared to YMS1\&2. 

 Panel (b) shows the spatial residual PM for YMS3. The SEA, SA, and WA are present in YMS3. However, the WA in YMS3 shows the northwest-directed motion of residuals, unlike the west-directed motion observed in YMS2. Additionally, the SA in YMS3 extends $\sim$ 6$^\circ$ further south than in YMS2. Also, the northern and southern outskirts of YMS3 exhibit higher residual PM magnitudes compared to YMS2. As a result, YMS3 shows a broader $|$residual PM$|$ distribution in panel (c), with an RMS value of 0.224$\pm$0.004 mas yr$^{-1}$.

\subsection{Clusters}
\begin{figure*}
    \centering
       \includegraphics[width=1\linewidth]{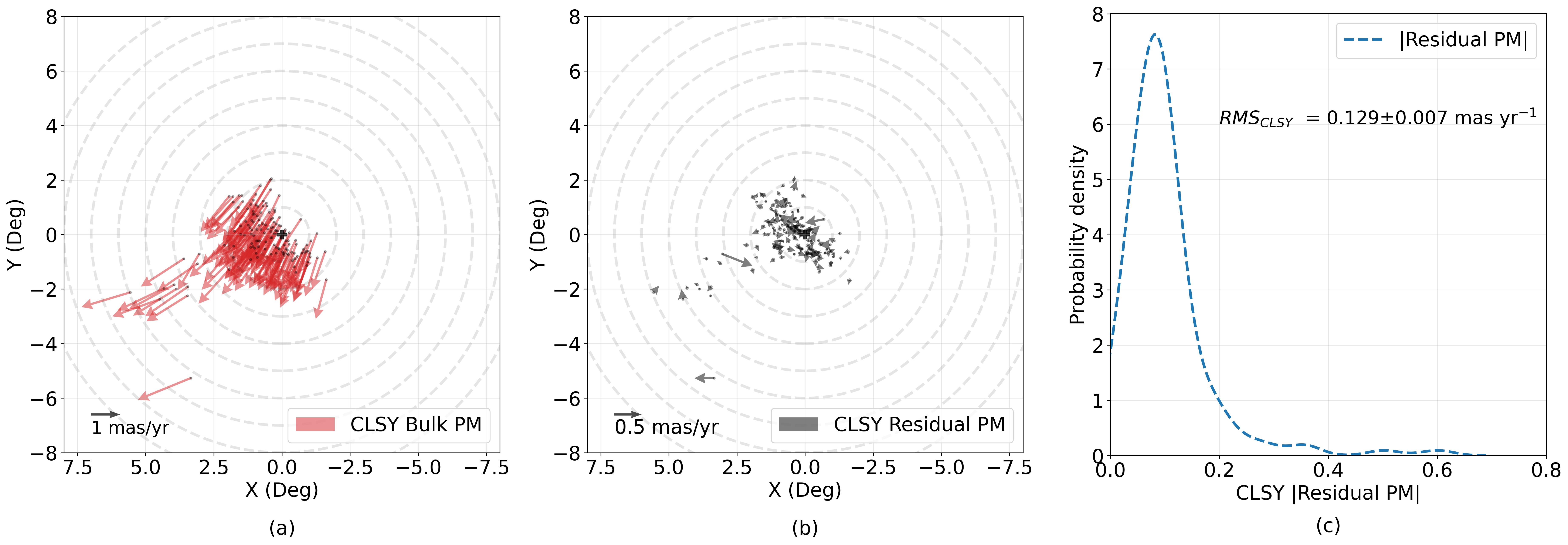}
       \caption{The observed PM (panel a), residual PM vectors (panel b), and distribution of $|$residual PM$|$ (panel c) for the CLSY are shown here.}
       \label{fig:c_pm} 
\end{figure*}

The clusters belonging to CLSY and CLSI show rotation signatures. We estimated $v_f$ and R$_f$ to be $\sim$ 67 km s$^{-1}$ and $\sim$ 7 kpc, respectively, for CLSY, and $\sim$ 50 km s$^{-1}$ and $\sim$ 6 kpc, respectively, for CLSI. This is consistent with the estimated kinematics of the YMS1 and YMS2. We also note the absence of significant rotation in the case of older clusters (CLSO with ages $>$ 1 Gyr), similar to the kinematics of YMS3. The COM PM does not show any significant deviation among the cluster groups as well as with respect to YMS1 and YMS2 populations. However, the estimated $i$ decreases from $\sim$ 80$^\circ$ for CLSY (young) to $\sim$ 63$^\circ$ for CLSO (old), meanwhile $\Theta$ increases from $\sim$ 180$^\circ$ to  $\sim$ 240$^\circ$, similar to the trend seen in the YMS population. 

Figure \ref{fig:c_pm}, panel (a) shows the observed bulk PM for the clusters belonging to the CLSY group. The clusters in our study are predominantly located in the eastern region of the SMC. Panels (b) and (c) show the spatial map and distribution of the magnitude of residual PM for CLSY. We note the residual PM vectors overall display a relatively more random orientation, along with several clusters showing a larger magnitude of residuals. The distribution of $|$residual PM$|$ for CLSY gives an RMS value of 0.129$\pm$0.007 mas yr$^{-1}$, which is less compared to YMS3, but more than that was found for YMS1\&2. Due to the sparseness, we do not detect the EA, SEA, SA, and WA in the cluster distribution. Also, we do not show the residual PM maps for CLSI and CLSO due to their sparse distribution in our sample.

\subsection{RGB and RC}
\begin{figure*}
    \centering
       \includegraphics[width=1\linewidth]{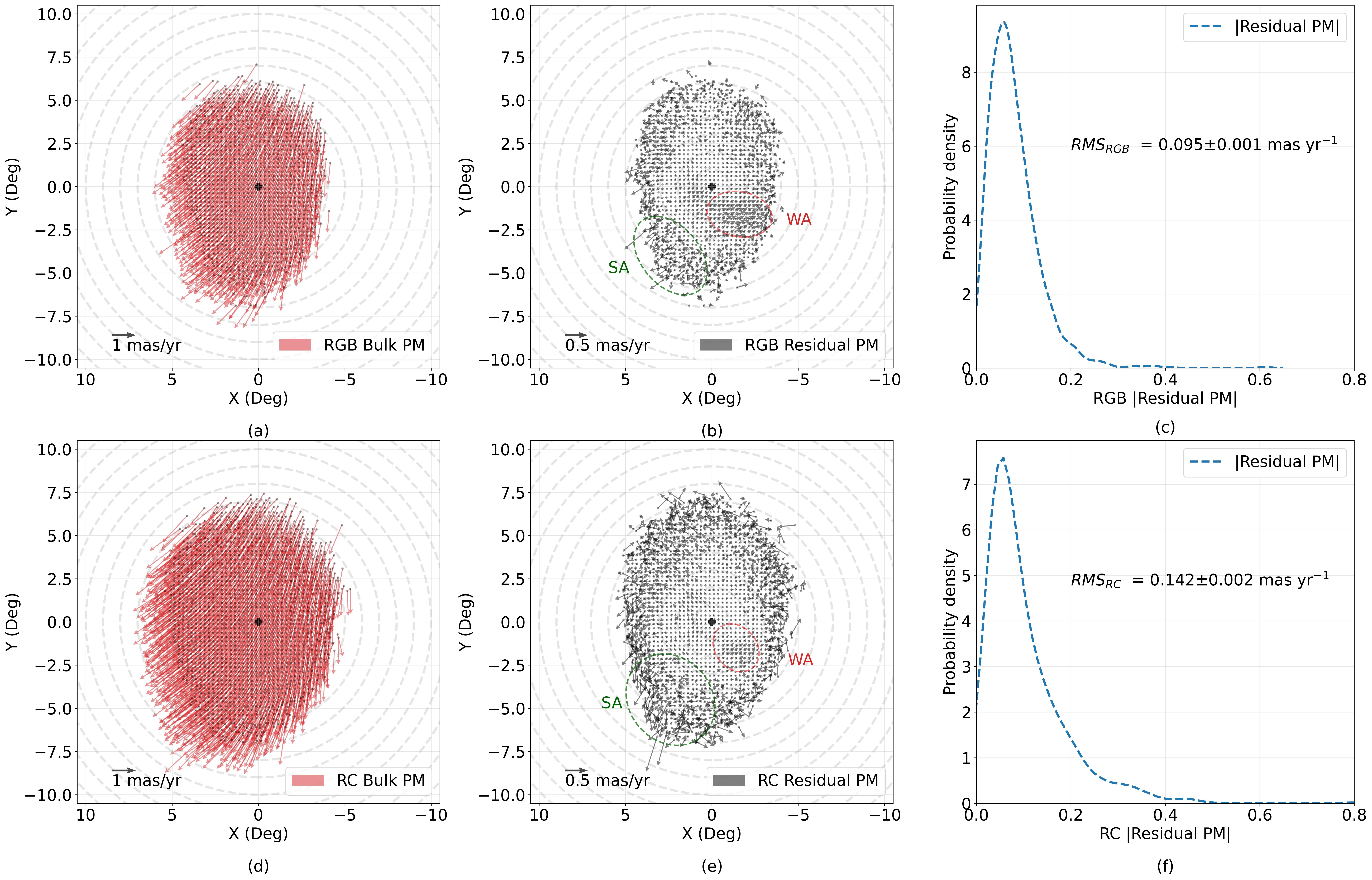}
       \caption{The observed PM (panels a and d), residual PM vectors (panels b and e), and distribution of $|$residual PM$|$ (panels c and f) for the RGB and RC are shown here. The SA and WA regions identified in the RGB and RC residual PM maps are highlighted.}
       \label{fig:rgb_rc_pm} 
\end{figure*}
Figure \ref{fig:rgb_rc_pm} presents the spatial map of the observed bulk PM for RGB and RC populations in panels (a) and (d), respectively. The spatial coverage of RGB and RC are similar, although RC extends slightly more towards the outskirts of the SMC. They do not support rotational components in their bulk PM. The estimated COM PM values for RGB and RC are comparable and do not exhibit significant deviations when compared to other populations, except for YMS3. However, the estimated $i$ is smaller, $\sim$ 58$^\circ$, while the $\Theta$ is estimated to be larger than 200$^\circ$.

Panels (b) and (e) in Figure \ref{fig:rgb_rc_pm} depict the spatial maps of residual PM vectors for the RGB and RC, respectively. The EA identified in YMS2 is observed beyond $\sim$ 3$^\circ$ and is slightly shifted northeast for both RGB and RC. Additionally, the SA anomaly extends beyond $\sim$ 4$^\circ$ to the south. However, the WA is more prominently evident for RGB than for RC. In the outskirts of the SMC, residual PMs are notably pronounced in the RC, especially in the northern and southern regions. 

The distributions of $|$residual PM$|$ for RGB and RC are shown in panels (c) and (f), respectively. The RC displays a higher RMS value of 0.142$\pm$0.002 mas yr$^{-1}$, compared to 0.095$\pm$0.001 mas yr$^{-1}$ for RGB.

\subsection{Red Giants with LOS velocities}

The estimated parameters for Red Giants with radial velocities (v$_{los}$) do not significantly deviate from the model parameters estimated for the RGB and RC populations without v$_{los}$. The $i$ ($\sim 66^\circ$) and $\Theta$( $\sim 191^\circ$) of the Red Giants are slightly deviant from the 2D model values for the RGB and RC, though these are still notably different from those of the younger main sequence population. Despite these differences, we do not observe any significant offset in the COM PM for Red Giants relative to the other populations. We estimated a systemic velocity (v$_{sys}$) of 146.70$^{\hspace{0.01cm}\scaleto{+0.10}{4.7pt}}_{\hspace{0.01cm}\scaleto{-0.10}{4.7pt}}$ for the Red Giants, which is closer to the assumed value of 145.6 km s$^{-1}$ used in the 2D modeling of other datasets. Since the radial coverage of Red Giants with v$_{los}$ is smaller ($\sim$ 3$^\circ$) and we modeled individual stars rather than binning them, we do not discuss the residual PM maps. 

The various populations in the SMC exhibit distinct kinematic properties, which aid in understanding the kinematic structure of the SMC based on age. In the following section, we discuss the intriguing details that emerge from our modeling.

\section{Discussion}\label{sec:disc}

In this study, we have carried out kinematic modeling of the SMC by analyzing nine different stellar populations, including YMS1, YMS2, YMS3, CLSY, CLSI, CLSO, RGB, RC, and Red Giants.

In the following subsections, we discuss the key results from our modeling. We compare the estimated kinematic parameters with those from previous studies, analyze the morphology of the galaxy, examine anomalies in the galaxy's internal motion, and explore the tidal evolution of the SMC. 

\subsection{kinematic parameters}
\begin{table}
    \centering
    \caption{Comparison of the estimated COM PM with previous studies. The first nine rows list the estimations from this study.}
    \label{tab_comparison_parameters}
    \renewcommand{\arraystretch}{1.5}
    \begin{tabular}{c@{\hspace{1.2em}}c@{\hspace{1.2em}}l}
    \hline
    \hline
        \textbf{$\mu_{W,com}$} & \textbf{$\mu_{N,com}$} & {Reference} \\
        (mas yr$^{-1}$) & (mas yr$^{-1}$) & \\
        \hline
        -0.657$\pm$0.008 & -1.262$\pm$0.007 & YMS1 \\   
        -0.712$\pm$0.010 & -1.283$\pm$0.010 & YMS2 \\
        -0.728$\pm$0.017 & -1.325$\pm$0.018 & YMS3 \\
        -0.704$\pm$0.004 & -1.252$\pm$0.005 & CLSY \\
        -0.687$\pm$0.005 & 
        -1.247$\pm$0.008 & CLSI \\
        -0.683$\pm$0.008 &
        -1.236$\pm$0.006 & CLSO \\
        -0.695$\pm$0.006 &
        -1.252$\pm$0.005 & RGB \\ 
        -0.687$\pm$0.008 & -1.249$\pm$0.008 & RC \\ 
        -0.687$\pm$0.008 & -1.249$\pm$0.008 & Red Giants \\ 
        -0.743$\pm$0.027 & -1.233$\pm$0.012 & N21 \\
        -0.721$\pm$0.024 & -1.222$\pm$0.018 & D20\\
        -0.82$\pm$0.10 & -1.21$\pm$0.03 & Z18\\
        -0.797$\pm$0.030 & -1.220$\pm$0.030 & G18\\
        -0.874$\pm$0.066 & -1.229$\pm$0.047 & V16\\
        -0.772$\pm$0.063 & -1.117$\pm$0.061 & K13\\
        -0.93$\pm$0.14 & -1.25$\pm$0.11 & C11 \\
        \hline
    \end{tabular} 
\end{table}

\begin{figure}
    \centering
       \includegraphics[width=1\linewidth]{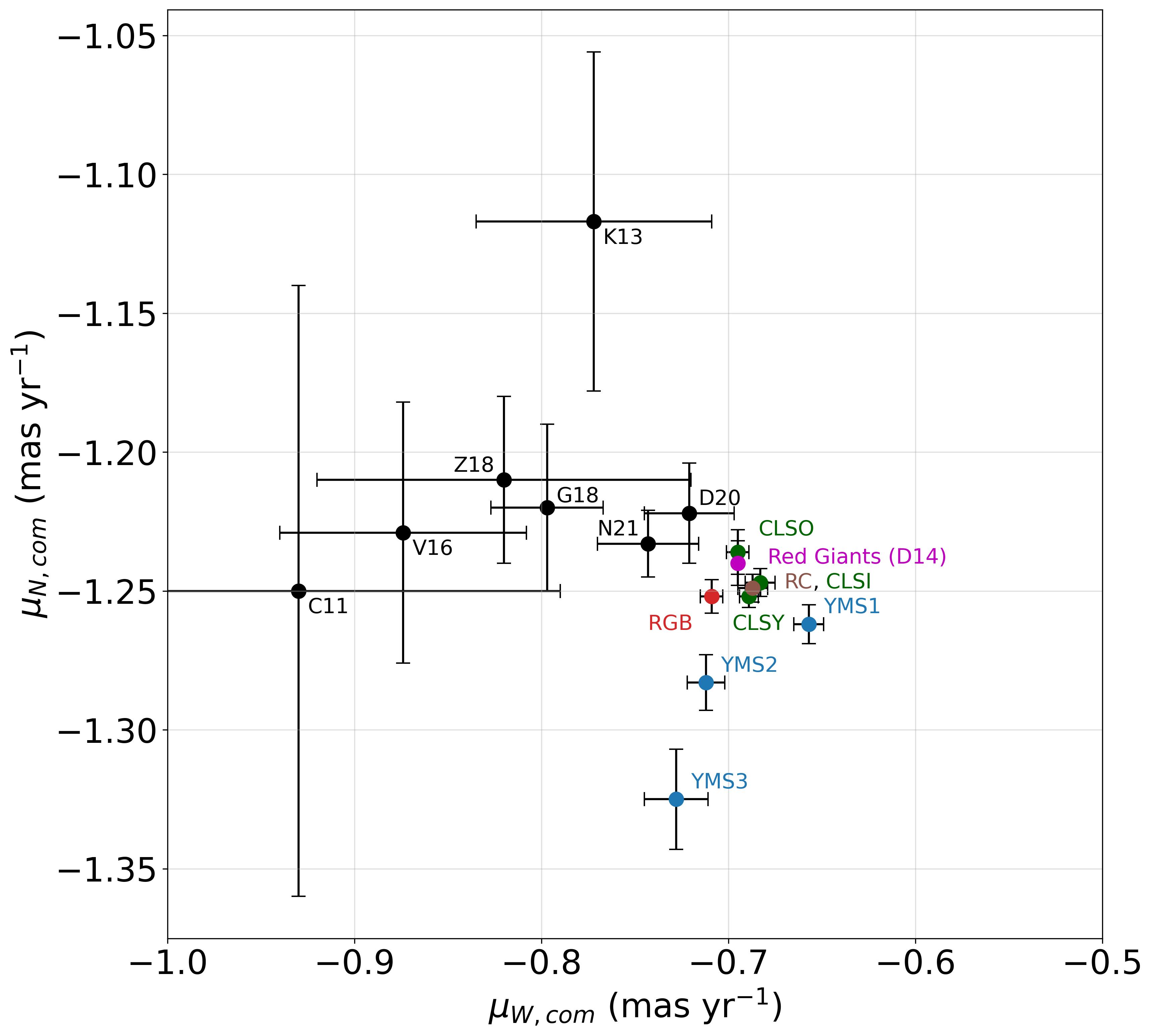}
       \caption{The parameter space of the estimated ($\mu_{W,com}$, $\mu_{N,com}$) is compared with the reference studies provided in Table \ref{tab_comparison_parameters}). The
corresponding estimated parameters for the YMS1, YMS2, and YMS3 are marked with blue dots. Clusters are marked with green dots. RGB, RC, and Red Giant red, brown, and magenta dots, respectively. The
reference studies are marked with black dots. Error bars are provided as well.}
       \label{fig:reference} 
\end{figure}

The estimated values of ($\mu_{W,com}$, $\mu_{N,com}$) of the SMC from our study are compared with those from previous studies by C11, K13, V16, Z18, G18, D20, and N21 in Table \ref{tab_comparison_parameters} and illustrated in Figure \ref{fig:reference}. The COM PM values for clusters (CLSY, CLSI, CLSO), RGB, RC, and Red Giants (with $v_{los}$) do not show significant offsets with respect to the recent estimates (D20 and N21). However, YMS1, YMS2, and YMS3 progressively exhibit offsets in the parameter space, with YMS3 displaying the largest offset.

 We note that the YMS3 population has ages between 1-2 Gyr, whereas YMS1\&2 are younger than 400 Myr. Also, the YMS3 shows a significantly high COM PM directed southward (see Table \ref{tab_comparison_parameters} and Figure \ref{fig:reference}). 
The largest southward shift seen in the YMS3 could be linked to the encounter between the LMC and the SMC about 1.5 Gyr ago (D24). This may be a sign that these stars probably retain the kinematic disturbance to the gas from which they are born, as a result of the interaction. The CLSO (ages $>$ 1 Gyr) does not show this trend since the clusters are smaller in number in this group.

The estimated value of $i$ for most of the old to young population in our study ranges from $\sim$ 58$^{\circ}$ to 82$^{\circ}$, meanwhile $\Theta$ ranges from $\sim$ 185$^{\circ}$ to 202$^{\circ}$. This range aligns with the findings of Z21 using \textit{Gaia} DR2 data, which reported a value of $i$ between 50$^{\circ}$ and 80$^{\circ}$ and a $\Theta$ of $\sim$ 180$^{\circ}$. S15 estimated ($i$, $\Theta$) of $\sim$ (64$^{\circ}$, 155$^{\circ}$) using cepheids. Meanwhile, D18 estimated these values to be $\sim$ (51$^{\circ}$, 66$^{\circ}$), using HI gas. The sparse distribution of CLSO, along with the central concentration of clusters in our sample results in a larger $\Theta$ value ($\sim$ 240$^\circ$). However, Our estimations tentatively suggest a decreasing value of  $i$ and an increasing value of $\Theta$ with the increase in age of the tracer population.

The bin sizes for binning the \textit{Gaia} data were set at 0$^\circ$.25 to visually identify the features seen in the PM and residual PM maps (Section \ref{sec:data}). Modifying the bin sizes, either by increasing or decreasing, as well as altering the minimum requirement of 5 stars per bin for estimating the median PM, does not substantially affect the parameter estimates in this study. We modified the bin sizes and the minimum number of stars per bin, but found that the parameter estimates do not change significantly across all the populations. Therefore, the estimated COM PM and viewing angles ($i$ and $\Theta$) for the SMC in our study are reliable and consistent with the findings of previous studies.

\subsection{Internal rotation of the SMC}\label{internal_discussion}

The younger populations show evidence of internal rotation in the SMC.
The value of $v_f$ we estimated in this study are $\sim$ 89 and 49 km s$^{-1}$ for YMS1 and YMS2, respectively. However, in our study, we estimated a larger R$_f$ of $\sim$ 9 and 6 kpc for YMS1 and YMS2, respectively, compared to the estimates by D18 (R$_f$  $\sim$ 1.1 kpc). A comparable trend is observed in the case of CLSY and CLSI. We were not able to fit the rotation component for the YMS3, CLSO, RGB, and RC, but we noted a partial convergence of v$_f$ below 8 km s$^{-1}$, which is not reliable as the error in the estimations is of the order of 10 km s$^{-1}$ (noted in YMS1\&2). This suggests that the rotation in the old population is significantly less, and can be assumed to have no rotation. This is consistent with the study by Z21, which found a moderate or slow rotation of $\sim$ 10 km s$^{-1}$ in the central regions of the SMC using Red Giant stars. 

\begin{figure*}
    \centering
       \includegraphics[width=1\linewidth]{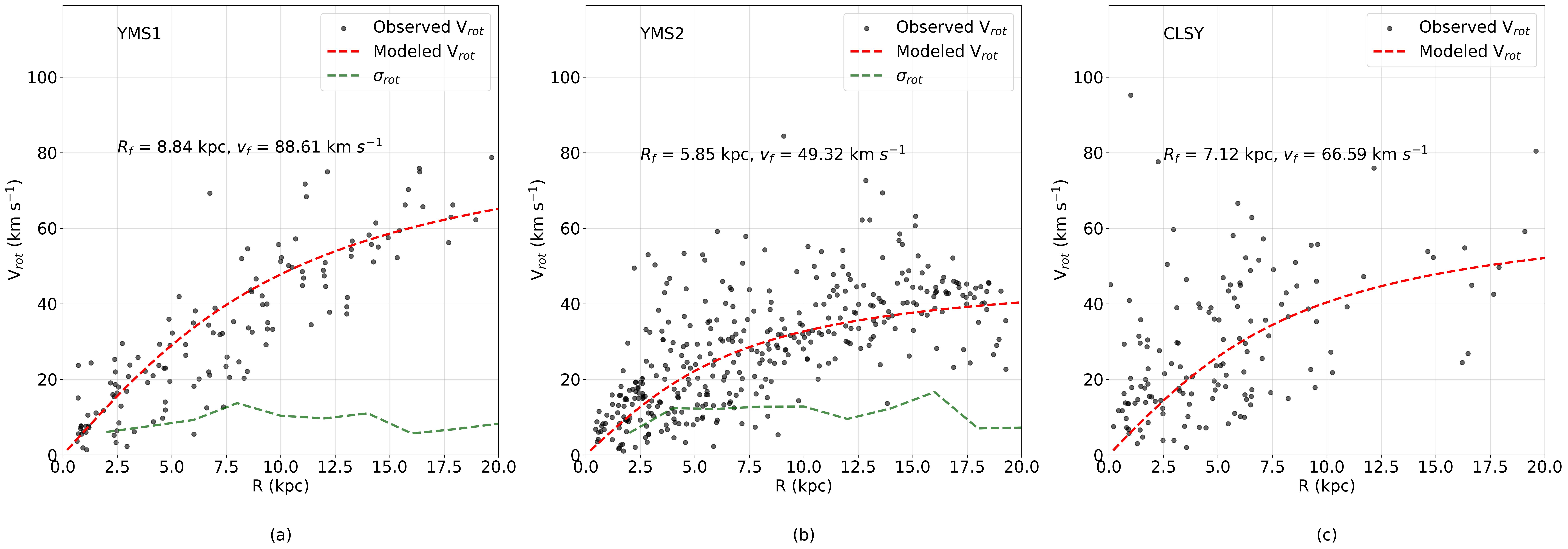}
       \caption{The rotation velocity (V$_{rot}$) profile of the SMC is shown for YMS1 (panel a), YMS2 (panel b), and CLSY (panel c) in the SMC plane. Black dots represent the observed V$_{rot}$, while the red dashed curve denotes the modeled V$_{rot}$. The dispersion ($\sigma_{rot}$) of the observed V$_{rot}$ for YMS1 and YMS2 populations are shown with the green dashed curve.}
       \label{fig:yms12rotation} 
\end{figure*}

Figure \ref{fig:yms12rotation}, panels (a) and (b) show the rotation velocity profile for the YMS1, YMS2, and CLSY respectively. The rotation velocity within the SMC plane (V$_{rot}$) is presented as a function of the galactocentric distance (R, in kpc). We also estimated an average rotational dispersion ($\sigma_{rot}$) of $\sim$ 9 km s$^{-1}$ and 11 km s$^{-1}$ for YMS1 and YMS2, respectively. This was done using a binning interval of 2 kpc along the disk radius. In CLSY, the absence of enough clusters beyond the radius of 11 kpc prevents the estimation of $\sigma_{rot}$. The smaller values of $\sigma_{rot}$ suggest a disk morphology for the younger population (YMS1\&2). The extension of R to $\sim$ 20 kpc or more indicates that the disk structure of the SMC is notably more stretched along the galaxy plane for the younger populations compared to its appearance in the sky plane.

For the cluster groups CLSY and CLSI, we found evidence of rotation consistent with the YMS1\&2 populations. However, the cluster samples in the wing region of the SMC are sparse in our study, leading to variations in $i$ and $\Theta$ between the young clusters (CLSY and CLSI) and the young main sequence populations (YMS1\&2).

In the next section, we present the in-plane morphology of the SMC based on our model.

\subsection{Morphology of the SMC}

\begin{figure*}
    \centering
       \includegraphics[width=1\linewidth]{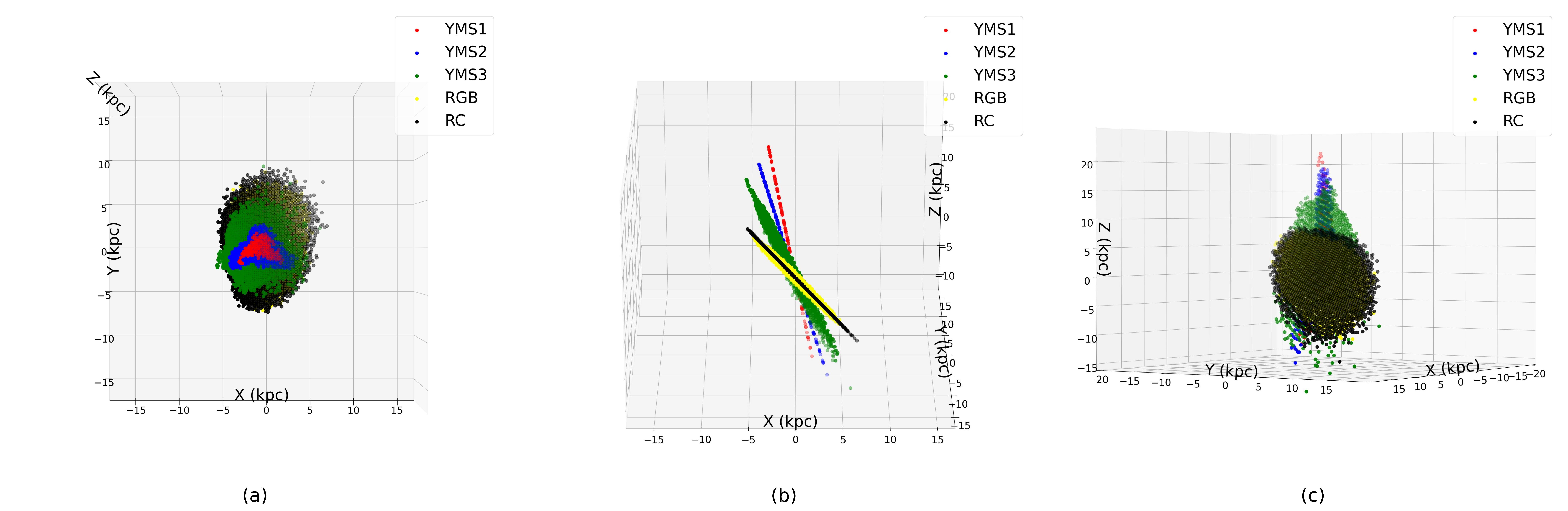}
       \caption{The distribution of YMS1, YMS2, YMS3, RGB, and RC populations in our study are color-coded and depicted here. (a) populations in the sky plane perspective (X-Y plane), (b) X-Y-Z perspective 1: ($R_1$, $R_2$) = (90$^{\circ}$, 20$^{\circ}$), (c) (b) X-Y-Z perspective 2: ($R_1$, $R_2$) = (90$^{\circ}$, 120$^{\circ}$). $R_1$ represents a clockwise rotation about the X-axis of the skyplane perspective, followed by $R_2$, which is a subsequent clockwise rotation about the new Y-axis.}
       \label{fig:smcxyz} 
\end{figure*}
Figure \ref{fig:smcxyz} shows the 3D perspective of the SMC for different populations (YMS1, YMS2, YMS3, RGB, and RC) based on the disk models we obtained in our study. The distances in X-Y-Z (in kpc) are estimated based on the fixed center ($\alpha_0$ = 13$^\circ$.05; $\delta_0$ = -72$^\circ$.83) and the mean distance (D$_0$ = 62.44 kpc) to the galaxy center. Panel (a) shows the density distributions of different populations as observed in the sky plane, where the negative X and Y directions are the East and South directions, respectively. Panel (b) shows the X-Y-Z perspective of the SMC, which is attained by the clockwise rotation of 90$^\circ$ about the X axis (Y=0), $R_1$ = 90$^\circ$ and followed by another clockwise rotation of 20$^\circ$ about the new Y axis, $R_2$ = 90$^\circ$. This keeps much of the population on the edge-on perspective, which also reflects their varying values of
$i$ for the disk. The RC and RGB disk has an offset of more than 19$^{\circ}$ with the YMS1\&2. Panel (c) shows an alternative X-Y-Z perspective of the SMC for $R_1$ = 90$^\circ$ and $R_2$ = 120$^\circ$, obtained after performing the rotations from the sky plane perspective from panel (a). The selection of $R_1$ and $R_2$ is arbitrary and it only serves to visualize the disk's inclination and the spatial distribution of different populations on the galaxy plane.

\begin{figure}
    \centering
       \includegraphics[width=1\linewidth]{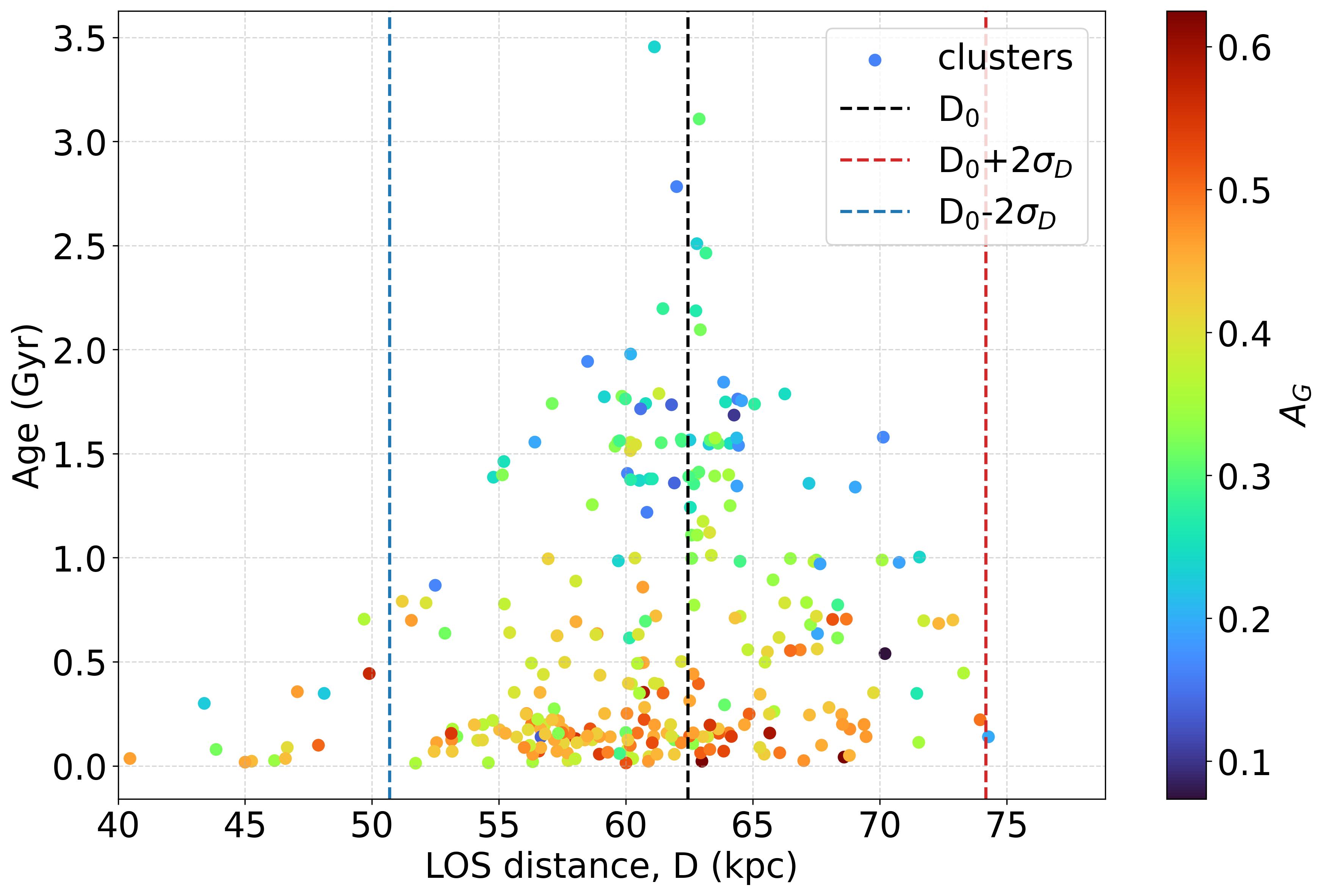}
       \caption{The age distribution of clusters (CLSY, CLI, and CLSO) is shown here according to their LOS distance (D), with each cluster color-coded based on extinction in the \textit{Gaia} G band (A$_G$). D$_0 = $ 62.44 kpc is the adopted mean distance to the SMC, with $\sigma_D$ representing the standard deviation in D based on the disk model.}
       \label{fig:D_clusters_age} 
\end{figure}

The triangular density distribution of the young population (YMS1\&2) in the sky plane appears similar to that of the observed density distribution of HI in the SMC (\citealt{h1_11981RMxAA...6...55L}, S04). The distribution of the older population (RGB and RC) in our study appears elliptical in the sky plane (Figure \ref{fig:smcxyz}, panel a). Also, the previous studies suggest that the old population of the SMC is distributed in a spheroidal or ellipsoidal shape (\citealt{Harris_2004}; R18; Y19). Our model indicates a highly inclined morphology for the SMC across all populations. We find that YMS1 and YMS2 are rotation-supported, while for the old population, we likely traced a projected geometry of ellipsoidal distribution along the semi-major and semi-minor axes. Notably, D14 estimated a LOS dispersion of $\sim$ 26 km s$^{-1}$ for the Red Giants we used in our study. This implies that the SMC probably has a flattened ellipsoidal distribution, where the younger populations (YMS1 and YMS2) have rotation-supported disk structures.

Panels (b) and (c) in Figure \ref{fig:smcxyz} reveal the LOS depth of the SMC based on the disk morphology obtained in this study for both young and old populations. Figure \ref{fig:D_clusters_age} shows the estimated LOS distance (D, in kpc) of clusters (CLSY, CLSI, and CLSO) in our study plotted against their age (in Gyr). The data is color-coded according to extinction in the \textit{Gaia} G band (A$_G$). The age and extinction values are taken from D24. We compare this plot with a similar plot by \citet[P21]{piatti202210.1093/mnras/stab3190} using clusters in the outer northeast of the SMC. We note a trail of young clusters ($<$ 200 Myr) in Figure \ref{fig:D_clusters_age} along the decreasing LOS distance (East of the SMC), which was noted in P21 as well. These clusters are also oriented along the wing region of the SMC, suggesting they were formed after the recent SMC-LMC interaction. The estimation of cluster ages in our previous study (D24) did not take disk morphology into account. Hence we re-estimated the ages of clusters using the current LOS distances obtained using the disk model of clusters. However, we observed minor differences in the estimates that fall within the margins of error for age estimation. The LOS depth for clusters, determined from the estimated inclination relative to the center, is $\sim$ 20 kpc to the east and $\sim$ 12 kpc to the west.

Our model suggests that the LOS depth for the RC and RGB populations extends $\sim$ 11 kpc to both the east and west of the SMC. In contrast, the LOS extension of the young populations (YMS1, YMS2, YMS3) in the west surpasses 15 kpc, with the eastern side of YMS1 extending to $\sim$ 20 kpc. The studies by J16 and R17 previously reported the younger population of the SMC to have LOS elongation of $\sim$ 20 kpc. In essence, our models trace that both the younger and older populations have a larger LOS depth, consistent with the previous observations. 

\begin{figure*}
    \centering
       \includegraphics[width=1\linewidth]{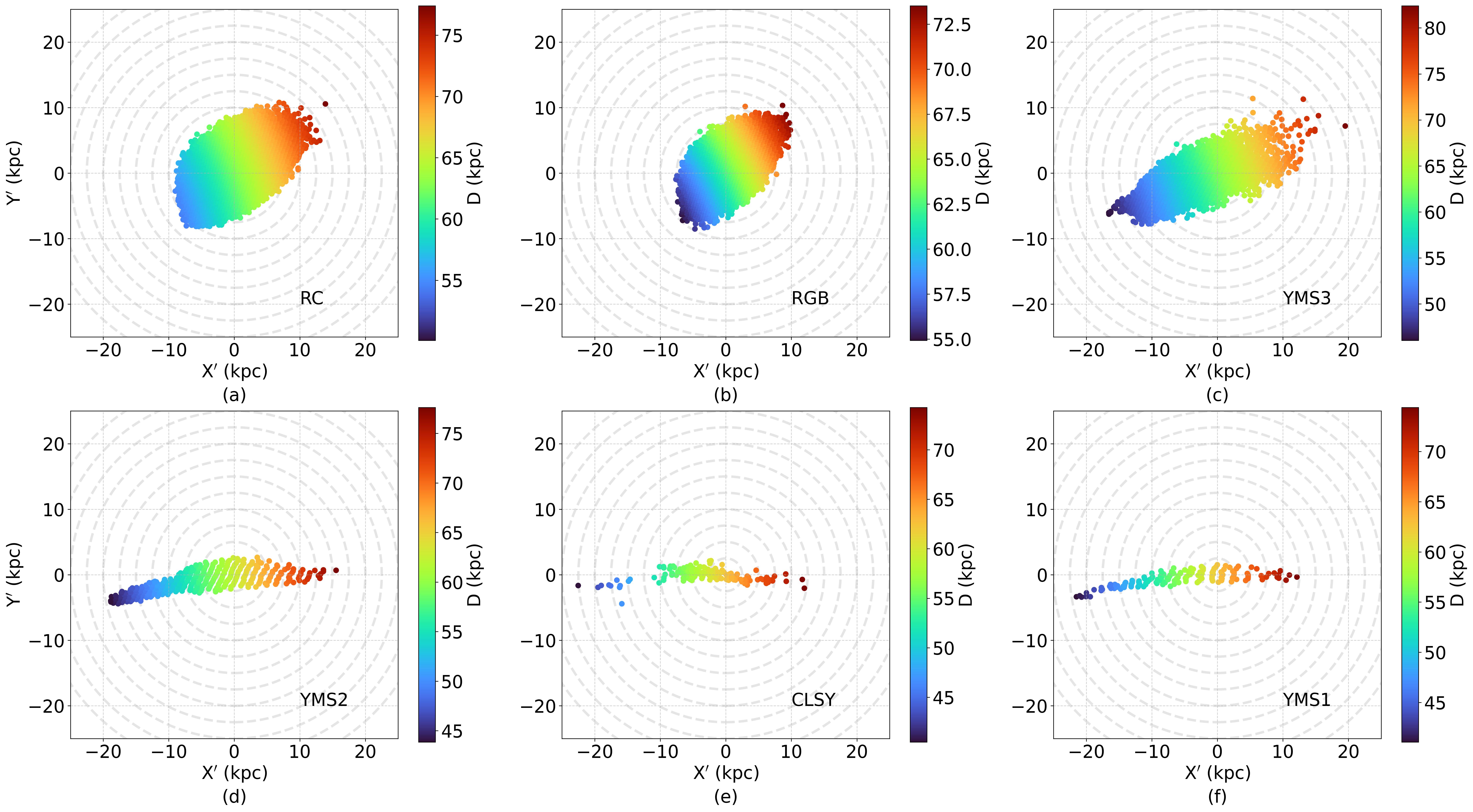}
       \caption{The morphology of the SMC is shown in the disk plane (primed coordinates) of the galaxy. Panels from (a) to (f) show RC, RGB, YMS3, YMS2, CLSY, and YMS1, respectively. The sources in each population are color-coded with LOS distance (D, in kpc) as well.}
       \label{fig:smcinplane} 
\end{figure*}
To better visualize the disk plane of the SMC, we present the RC, RGB, YMS3, YMS2, CLSY, and YMS1 within their best-fitted disk plane of the galaxy (X$^{\prime}$-Y$^{\prime}$ axes), as shown in Figure \ref{fig:smcinplane}. Each population's disk plane is rotated counterclockwise by their corresponding $i$ and about the PA of the line of node axis. Also, each source is color-coded based on the LOS distance (D, in kpc) as well. This perspective aligns the disk plane with the standard East, West, North, and South directions of the sky plane, facilitating a comparison of the disk plane and its appearance in the sky. Starting with the RC and RGB populations, we observe that the SMC extends up to $\sim$ 10 kpc from its center. The YMS3 population, which is 1-2 Gyr old, is more spread out, but elongated compared to the RC and RGB, reaching over $\sim$ 17 kpc from the SMC center. In contrast, the YMS1, YMS2, and CLSY, which are less than 400 Myr old, exhibit an even larger stretch in their disk distribution. These populations are significantly more elongated along the X$^{\prime}$ axis compared to the Y$^{\prime}$ axis, extending up to $\sim$ 20 kpc from the center, with YMS1 and CLSY spanning more than 22 kpc. However, we estimated a small dispersion of $\sim$ 10 km s$^{-1}$ in the rotation profiles for the YMS1 \& 2 (Figure \ref{fig:yms12rotation}, panels a and b). This suggests that the young SMC likely has a disk structure that is not very thick. Additionally, the appearance of the RC and RGB indicates a more compact ellipsoidal structure compared to the young populations.

In the following section, we will investigate the anomalies identified in the residual PM maps in conjunction with the SMC's morphology.

\subsection{Anomalies in the kinematics and tidal evolution of the SMC}\label{last_section}

The anomalies we identified in the residual PM map of YMS2 (Section \ref{sec:results}) are evident in both the young and the old populations. The SEA, which represents a stellar population in counter rotation is noticed for the first time in the SMC, and it is prominent across the YMS1, YMS2, and YMS3. The appearance of counter-rotation in the residual could arise due to the slower rotation of the population with respect to the model (This is explored further in detail later). The EA indicates the motion of groups of stars away from the SMC and directed to the LMC along the young bridge of the SMC. We note the clusters in the east wing are mostly of ages less than 200 Myr, they also have residual PM aligned towards the younger bridge. However, the SA shows the residual PM vectors of the stellar population aligning to the south in the SMC outskirts, probably connecting to the old bridge.

The eastward movement of stars in the younger population toward the younger bridge was previously noted in the study by N21 and aligns with our findings. Figure 16 of G21 illustrates comparable movements of both old and young populations toward the older and younger bridges, respectively, which is consistent with our findings. The residual PM features located beyond $\sim$ 5$^\circ$ to $\sim$ 8$^\circ$  for RGB and RC (as shown in Figure \ref{fig:rgb_rc_pm}, panels b and e) show similarities to the trends observed by \cite{lara2023MNRAS.518L..25C} for the offset population of RGB stars in the SMC outskirts, as depicted in their Figure 4. 

The WA is observed in the central regions and extends mostly westward in the SMC, which is evident in both young and old populations (YMS2, YMS3, RGB, and RC). The west-directed motion of these populations is consistent with the studies by \cite{nid2018A&A...613L...8N} and \cite{vis2022MNRAS.512.4334D}. The motion is directed to the west halo of the SMC, a tidal feature suggested by \cite{diaz2016A&A...591A..11D}. The west halo is later associated with the counter bridge of the SMC \citep{tatton2021MNRAS.504.2983T}, which loops behind the SMC from southwest to northeast of the galaxy.  

\begin{figure*}
    \centering
       \includegraphics[width=1\linewidth]{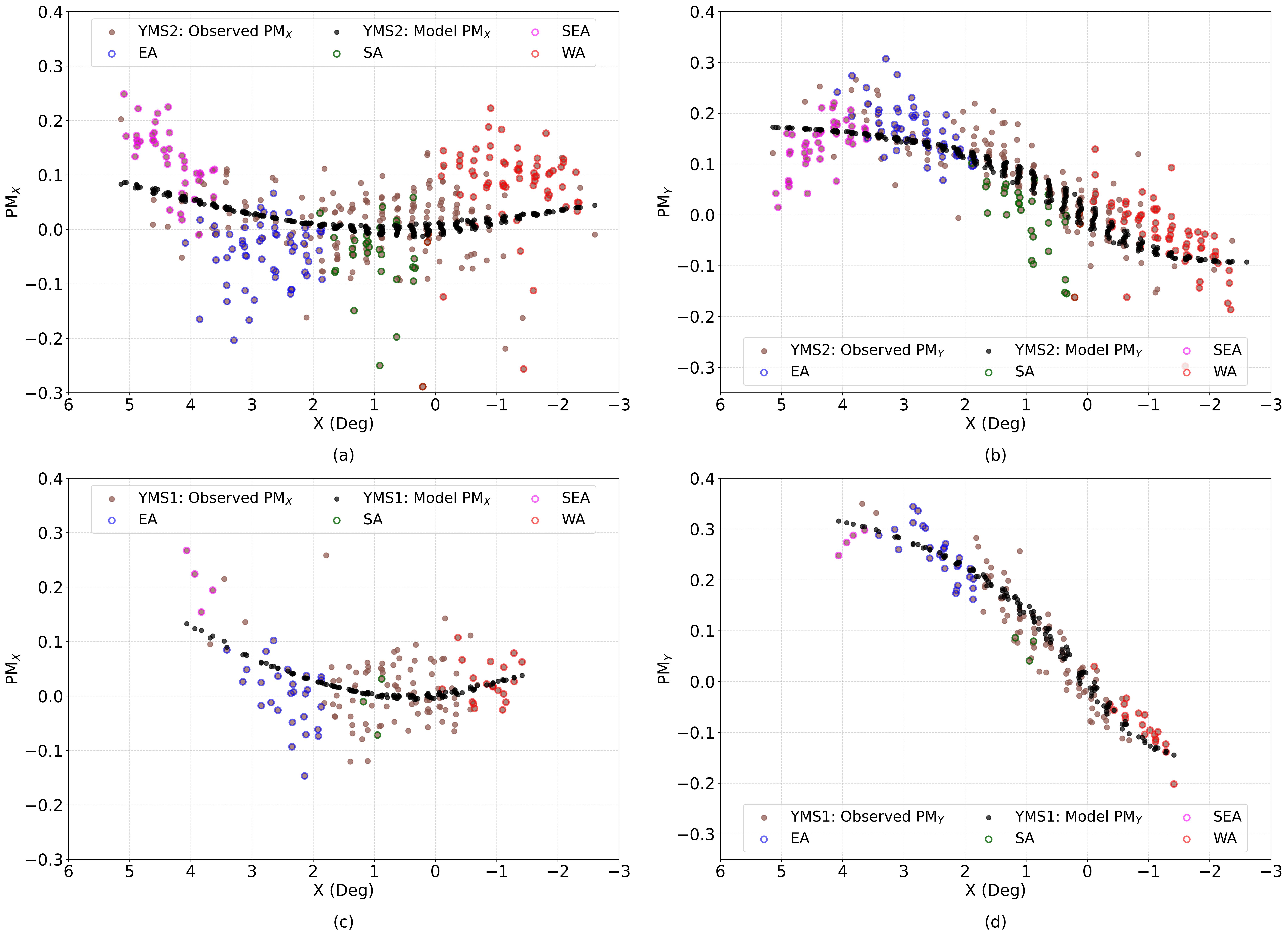}
       \caption{Comparison of the internal PM components (PM$_X$, PM$_Y$) along the X direction of the sky plane for YMS2\&1 are shown here. Observed PM$_X$ and PM$_Y$ are represented by brown dots, while their corresponding model predictions are indicated by black dots. The SEA, EA, SA, and WA binned stars identified for YMS2 (see subsection \ref{yms1_2_results}) are marked with magenta, blue, green, and red circles, respectively.}
       \label{fig:tangential_aniso} 
\end{figure*}

The anomalies observed point out that the galaxy is showing tangential anisotropy in PM across all the populations. We further analyze the observed internal PM$_X$ and PM$_Y$ along the X direction for YMS2\&1, aiming to assess their respective contributions to these four anomalies. In Figure \ref{fig:tangential_aniso}, panels (a) and (c) illustrate the variation in the X component of the internal PM (PM$_X$) for YMS2 and YMS1 along the X direction in the skyplane. Meanwhile, panels (b) and (d) depict the variation in the Y component of the internal PM (PM$_Y$) for YMS2 and YMS1 along the same X direction in the skyplane. The color-coding for anomalies in both YMS1 and YMS2 across each panel is based on the EA, SEA, SA, and WA regions identified in YMS2. This facilitates a region-to-region comparison between YMS2 and YMS1.

The SEA is found to have larger values of (PM$_X$) indicative of an inward motion, and lower values for (PM$_Y$) indicative of slower rotation, particularly for regions at X $\sim$ 4$^\circ$ and beyond as seen in panels (a) and (b). Furthermore, the SEA region in YMS1, as shown in panels (c) and (d), noticeably has a reduced number of bins compared to YMS2. This suggests that the disk's rotation is attaining stability in the youngest population (YMS1). The SEA prominently emerges from the YMS3 population (ages $\sim$ 1 to 2 Gyr) and is not so evident in RGB and RC. This suggests that the SEA is predominantly detected in the populations with $\sim$ 50 Myr to 2 Gyr age range. Our models indicate that the SMC's rotation starts with the YMS2 population, suggesting that the SEA may be pointing to a gas in-fall from the outskirts into the southeastern region of the SMC, where stars formed from this gas retaining the in-falling gas kinematics.  We therefore classify the SEA to be an infalling feature, rather than a counter-rotating feature.

The EA reveals larger negative values in PM$_X$, and it is evident in both YMS2 and YMS1. The inner bins in the east between $\sim$ 2$^\circ$ and 4$^\circ$ of YMS2 show large negative observed PM$_X$ values, suggesting the eastward directed motion. This implies that stars in these bins are being pulled eastward towards the LMC. In the YMS1 population, the bins of stars in the central regions between $1^\circ$ and 3$^\circ$ exhibit similar eastward motion, along with larger negative values for PM$_X$. This points to the tidal signature in the RA direction likely resulting from the SMC interacting with the LMC. 

The SA reveals the flow of stars towards the south of the SMC, which is prominent in the YMS2 population, and only marginally present in the YMS1. The outward motion of binned stars to the south is also observed in the YMS3, RGB, and RC populations, suggesting that this is likely associated with the old bridge of the SMC. 

The WA appears prominently in YMS2 but not in YMS1. The more positive values of PM$_X$ observed for WA when compared to the model, are suggestive of a westward motion. There is a marginally higher value for these populations in the PM$_Y$ when compared to the model suggesting a northward motion. A predominantly westward motion of stars present in YMS2 is missing in YMS1. If the gas in these regions is also affected by this motion, then the gas can escape the SMC through the western halo. This could explain the contribution of gas to the Magellanic stream which in turn could explain the absence of stars in YMS1 for values of X $<$ -1$^\circ$ (panel c). Notably, all the anomalies (SEA, EA, SA, and WA) show greater anisotropy in PM$_X$ than in PM$_Y$, which is a result of the line of interaction with the LMC. Additionally, the variation of PM$_X$ and PM$_Y$ along the Y direction shows similar anomalies.

The morphology of the young SMC, as seen in the SMC plane, appears to be extensively stretched in the east-west direction. This stretching is starting to appear in the YMS3 population and is full-blown in the distribution of YMS2 population. The contrasting distribution of the RC/RGB (panel a/b) to the YMS2 (panel d) in Figure \ref{fig:smcinplane} highlights the skewed stellar distribution. The morphological change is suggestive of a stretched distribution of the gas that is formed in the younger populations. 
We explore the possible connection of this feature with the tidal effects of the LMC-SMC interaction. The eastward motion is likely to be associated with the recent interactions between the LMC and the SMC, as evidenced by the eastward stretch across all populations. Similarly, the westward motion is evident across all populations, but less pronounced in the YMS1. These two interaction signatures, which are acting along the same line, but in opposite directions in the SMC's internal PM suggest that its elongation is a result of the significant tidal stretching, particularly from the recent interaction, as it is most pronounced in YMS2. Moreover, the increased inclination of the younger population has resulted in smaller LOS distances in the eastern part, along with a significant reduction in the extent of the SMC in the north-south direction. All these point to a significant tidal influence from its larger companion. The fact that all anomalies exhibit more deviation in the X-direction of the PM is also suggestive of extensive tidal signatures across the SMC. The tangential anisotropy on the PM of the SMC has previously been observed in the RA direction of the galaxy, which is associated with the tidal disruption of the dwarf galaxy due to the gravitational influence of the LMC (\citealt{klimentowski10.1111/j.1365-2966.2007.11799.x}; D18; \citealt{hota2024uvitstudymagellanicclouds}). Similarly, several studies have reported tidal stripping or stretching of the SMC (\citealt{piatti2021A&A...650A..52P}b, N21, \citealt{yous2023MNRAS.523..347E}), which aligns with the picture of the SMC emerging from our models. The counter-rotating signature that is identified in this study for the first time is unlikely to be a tidal signature, but a hydrodynamic one probably from the gas-infall in the course of the morphological evolution of the SMC.

This study provides baseline models estimated from the median motion as traced by the majority of stellar populations across the SMC. The anomalies detected in the residual PM of the different stellar populations are likely to be the major outliers as they are shown by the bulk of the population. This study therefore does not detect kinematic and structural outliers that exist in smaller fractions of the population across the SMC. Therefore, we note that the base models derived here will be very helpful in detecting kinematic and structural outlier (minority) populations that exist in the SMC. The unique PM pattern identified in this study can aid numerical simulations to pinpoint the details of the interaction between the MCs. However, the anomalies seen in the residual PM maps of this study need further investigation, such as a 3D model including radial velocity for all populations to fully comprehend the dynamic evolution of the SMC.

\section{Summary}\label{sec:summary}
We summarize the results and conclusions derived from the disk model of the SMC using Gaia DR3 data.
\begin{itemize}
    \item We performed the kinematic model of the SMC using nine different populations to investigate the galaxy's morphology. Eight populations were used to derive a 2D model using PM from \textit{Gaia} DR3 and one was used to derive a 3D model using PM (\textit{Gaia} DR3) and radial velocities. The data coverage of the SMC considered in this study is within $\sim$ 7$^\circ$.5 from the galaxy center. 
    \item The best fitting kinematic parameters of the SMC were estimated using a Markov Chain Monte Carlo (MCMC) method. The parameters estimated in this study, such as the inclination of the SMC disk ($i$), the position angle (PA) of the line of nodes measured from West ($\theta$), the amplitude of the tangential velocity of the SMC's center of mass ($v_t$), the tangential angle made by $v_t$ ($\theta_t$), scale radius ($R_f$), the asymptotic velocity ($v_f$), and the systemic velocity ($v_{sys}$)  show good agreement with estimations in the literature when comparisons are made between similar populations. 
    \item The COM PM for the entire population exhibits minimal variation, except in YMS3, where the southward motion is larger than other populations. This may be a signature of the LMC-SMC interaction $\sim$ 1-2 Gyr ago.
    \item We estimated the $i$ to range from $\sim$ 58$^\circ$ to 82$^\circ$, and $\Theta$ range from $\sim$ 180$^\circ$ to 240$^\circ$ among the young and old population of the SMC. We observe that $i$ decreases with age, while $\Theta$ increases.
    \item We estimated a v$_f$ of $\sim$ 89 km s$^{-1}$ for YMS1 and 49 km s$^{-1}$ for YMS2, with corresponding R$_f$ of $\sim$ 9 kpc and 6 kpc. A similar trend is observed for CLSY and CLSI. This suggests that both YMS1 and YMS2 show a rotation-supported disk structure for ages less than 400 Myr. In contrast, the older populations (YMS3, RGB, RC, CLSO, and Red Giants) do not exhibit significant rotation and are likely pressure-supported.
    \item The young main sequence population (YMS1 and YMS2) shows an elongated structure in the galaxy plane, with a rotational velocity dispersion of $\sim$ 11 km s$^{-1}$, suggesting a flattened rotating structure for the SMC.
    \item Our models reveal a larger LOS extension for the SMC reaching up to $\sim$ 30 kpc across the different stellar populations (old and young).
    \item We identified several anomalies on the residual PM of YMS2, which are the East Anomaly (EA), South East Anomaly (SEA), South Anomaly (SA), and West Anomaly (WA). The SEA identified for the first time is suggestive of a infalling population possibly having a hydrodynamic origin. The SA is likely associated with the old bridge, the EA and WA appear to be of tidal origin.
    \item This study also brings out the morphological change in the SMC over its evolution. The extensive east-west stretch seen in the young population is likely to be due to the skewed distribution of gas in the SMC resulting from the recent interactions. 
    \item This preferential stretching in the X-direction is also noted in the young stars. The internal rotation of the young population along the X-direction (PM$_X$) of the SMC exhibits greater tangential anisotropy than in the Y-direction, suggesting that the galaxy is being tidally stretched due to the influence of the LMC. 
\end{itemize}
\section{Acknowledgements}
We thank the referee for valuable suggestions that helped to improve the manuscript. Annapurni Subramaniam acknowledges support from the Science and Engineering Research Board (SERB) Power fellowship. This work made use of the optical data from the European Space Agency (ESA) mission \textit{Gaia} (\url{https://www.cosmos.esa.int/gaia}), which was processed by the \textit{Gaia} Data Processing and Analysis Consortium (DPAC,
\url{https://www.cosmos.esa.int/web/gaia/dpac/consortium}). The funding for the DPAC has been provided by national institutions, particularly the institutions participating in the \textit{Gaia} Multilateral Agreement. 

{\it Software:} ASTROPY \citep{astro_1_2013A&A...558A..33A,astro2_2018AJ....156..123A,astro3_2022ApJ...935..167A}, SCIPY \citep{2020SciPy-NMeth}, MATPLOTLIB \citep{matplot2007CSE.....9...90H}, NUMPY \citep{numpy2020Natur.585..357H}, CORNER \citep{corner} 
\bibliography{sample631}{}
\bibliographystyle{aasjournal}

\end{document}